\documentclass[aps,prl,groupedaddress,twocolumn]{revtex4}%
\pdfoutput=1
\usepackage{dsfont}
\usepackage{epsfig, amssymb,amsthm}
\usepackage{bbm}
\usepackage{times}
\usepackage{amsmath}
\usepackage{color}%
\usepackage{amsfonts}%
\usepackage{amssymb}%
\usepackage{graphicx}

\providecommand{\U}[1]{\protect\rule{.1in}{.1in}}

\definecolor{red}{rgb}{0.99,0.40,0.0}

\newcommand{\bra}[1]{\langle#1|}
\newcommand{\ket}[1]{|#1\rangle}

\begin{document}
\title{Measuring Measurement:  Theory and Practice}
\author{A.\ Feito$^{2,3}$}
\email{ab1805@imperial.ac.uk}
\author{J.~S.\ Lundeen$^{1}$}
\author{H.\ Coldenstrodt-Ronge$^{1}$}
\author{J.\ Eisert$^{4}$}
\author{M.~B.\ Plenio$^{2,3}$}
\author{I.~A.\ Walmsley$^{1}$}
\affiliation{$^{1}$Clarendon Laboratory, Oxford University, Parks Road, Oxford, OX1 3PU, UK}
\affiliation{$^{2}$Institute for Mathematical Sciences, Imperial College London, SW7 2PG,
UK }
\affiliation{$^{3}$QOLS, The Blackett Laboratory, Imperial College London, Prince Consort
Road, SW7 2BW, UK}
\affiliation{$^{4}$Physics and Astronomy, University of Potsdam, 14476 Potsdam, Germany }
\date{\today}

\begin{abstract}
Recent efforts have applied quantum tomography techniques to the calibration
and characterization of complex quantum detectors using minimal assumptions.
In this work we provide detail and insight concerning the formalism, the
experimental and theoretical challenges and the scope of these tomographical
tools. Our focus is on the detection of photons with avalanche photodiodes and
photon number resolving detectors and our approach is to fully characterize
the quantum operators describing these detectors with a minimal set of well specified
assumptions. The formalism is completely general and can be applied to a
wide range of detectors.

\end{abstract}
\maketitle



\section{Introduction}

The quantum properties of nature reveal themselves only to carefully designed
measurement techniques \cite{Resch, Pryde}. In addition, most quantum
information applications both computational and cryptographic, rely on a
certain knowledge of the measurement apparatuses involved. Indeed for these
protocols we often need to ensure that we associate detector outcomes with the
correct quantum mechanical operation or quantum state. More critically,
the assumption of a fully characterized detector completely underlies both
quantum state tomography (QST) and quantum process tomography (QPT)
\cite{PhysRevLett.70.1244,PhysRevLett.78.390,kwiat}. State tomography has
become an important tool for characterizing states, partially due to the
realization that non-classical states are a resource for performing tasks such
as enhanced precision metrology, quantum communication, and quantum
computation. Often taken for granted, measurement also plays a crucial role in
these tasks and in some can even eliminate the need for entanglement. In
enhanced precision metrology, appropriate measurements alone can give rise to
super-resolution \cite{Resch} and Heisenberg-limited sensitivity \cite{Pryde}.
In communication, measurement allows entanglement swapping, and thus, is
central to quantum repeaters. And for computing, measurement based schemes
enable quantum computation \cite{Raussendorf, Raussendorf-Briegel-prl}. It
follows that measurement should also be considered a resource for quantum
protocols. In QST a given number of measurements on many copies of an unknown
state reveal its density operator \cite{PhysRevA.40.2847, PhysRevLett.70.1244,
PhysRevA.60.674}. Characterising the operators that govern an evolution or a
channel -- QPT -- amounts to applying the process to a set of input states,
and subsequently fully characterising the output states \cite{chuang-1996,
PhysRevLett.78.390,PhysRevLett.80.5465,kwiat,lvovsky-2008}. In this
paper we study quantum detector tomography (QDT) \cite{PRLSoto,
dariano-2004-93, PhysRevA.64.024102,JMOTomo, NaturePhysTomo}, in which a
detector's outcome statistics in response to a set of input states determine
the operators that describe that detector. State and detector tomography
evidently exhibit a dual role: Either the input is well-known and the detector
is to be characterised, or the detector is well-known and the state is to be
tomographically reconstructed.\\

Throughout the paper, we focus on examples from optics. However, quantum
detector tomography is a general concept, applicable to any quantum detector.
Building upon previous theoretical descriptions \cite{PRLSoto,
dariano-2004-93, PhysRevA.64.024102,JMOTomo} we will introduce the concept of
detector tomography in the next section. Alongside, we will present
examples detailing the reconstruction of simple detectors to introduce the key
concepts. Subsequently, based on recent experiments \cite{NaturePhysTomo} we
will present the methods used in optical detector tomography and the convex
optimization methods \cite{Convex} which allow an efficient and simple
numerical optimization. Finally, in the last section, we will discuss some of
the theoretical and experimental challenges involved and how to address them.

\section{Detector tomography}

\subsection{Definitions}

In quantum mechanics, the operator describing a measurement apparatus is, in
its most general form, a positive operator valued measure (POVM). The
POVM elements $\{\pi_{n}\}$ describe the possible outcomes, labeled here by
$n$. Particularly, for a projective measurement, the POVM elements are
orthogonal and simplify to the familiar form
\[
\pi_{n}=|\psi_{n}\rangle\langle\psi_{n}|.
\]
In quantum optics, an example of a POVM that consists of projectors is
that of an eight-port homodyne, $\{|\alpha\rangle\langle\alpha|:\alpha
\in\mathbbm{C}\}$.

Now, given an input state $\rho$, the probability $p_{\rho,n}$ of obtaining
output $n$ is%
\begin{equation}
p_{\rho,n}=\mathop{\rm Tr}{_{}}\left(  \rho\pi_{n}\right)  .\label{Trace}%
\end{equation}
It follows that the POVM elements must be positive semi-definite, $\pi_{n}%
\geq0$, while observing
\begin{equation}
\sum_{n}\pi_{n}=\mathbbm{1},
\end{equation}
if we want probabilities adding up to one. Inverting Eq. (\ref{Trace}) to
extract $\pi_{n}$ subject to the aforementioned conditions is the task behind
detector tomography.

\subsection{Assumptions}

Detector tomography raises fundamental questions about the kind of information
we can extract from Nature. It is reasonable to think that state tomography
performed with poorly characterized detectors can lead to unwanted errors. In
addition, for quantum cryptography, a mischaracterization of states or
detectors may lead to channels through which an eavesdropper may attack
invalidating certain security proofs (see for ex. \cite{Makarov}).  Indeed,
if we misjudge the noise level, then we misjudge the information the eavesdropper possesses.
Interestingly, quantum key
distribution with totally uncharacterized detectors is possible in principle,
based entirely on correlations violating Bell's inequalities, but at the
expense of having much smaller rates \cite{Masanes1,Masanes2}. 
An in depth discussion of our assumptions is therefore
necessary to avoid unwanted errors in our detector estimation. 

Generally, there are going to be assumptions about the input states
produced by the source.  In the reconstruction, we will often need to
assume we can truncate an infinite dimensional Hilbert space, such as
in the case of photon number. On the detector side, a common
assumption will be that the detector is memory-less: the previous
measurements do not modify the result of future measurements.  For
example, this assumption fails when detector deadtimes are
longer than the time between consecutive measurements.  All of these
assumptions need to be tested.

More generally, we can ask what the working minimal set
of assumptions happens to be. An assumption-free tomography could use a
complete black box approach: prepare a collection of unknown states, measure
them and try to draw some conclusions about both the detector and the states.
Specifically, we could have some classical controls to prepare quantum states
$\{\rho_{\lambda}\}$ characterized by the index $\lambda$ and some classical
pointer to indicate the possible outcomes, labeled $n$. Minimizing the set of
assumptions would constrain us to draw our conclusions exclusively from the
joint probability distribution $\{p_{\lambda,n}\}$. \newline\newline To
further constrain the problem we can add the standard assumptions: An
underlying Hilbert space of fixed dimension, normalized positive density
matrices and positive POVM elements $\{\pi_n\}$ satisfying $\sum_{n=1}^{N}%
\pi_{n}=\mathbbm{1}$. However, without further assumptions, the relationship
between $\lambda$ and $\rho$'s $d^{2}-1$ parameters is completely unknown, as
is that between $n$ and $\{\pi_{n}\}$'s $(N-1)d^{2}$ parameters.

This discussion highlights the inherent difficulties that a fully general
inference (or tomographic) scheme entails. Reasonable assumptions are thus
needed but the question of general tomography remains an interesting one to be
explored. In this direction, some progress has been made in self-testing maps.
In this context states are prepared with classical recipes and families of
unitary gates are revealed with few assumptions about the quantum states
(however known measurements in the computational basis are required)
\cite{self-testing}.
\begin{figure}[ptbh]
\begin{center}
\includegraphics[width=6.5cm]{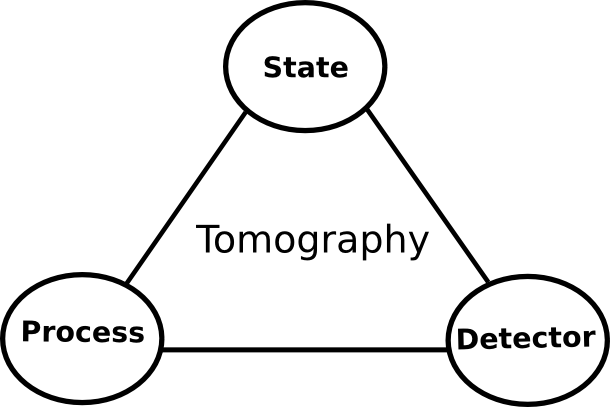}
\end{center}
\caption{It is generally possible to divide an experiment into preparation,
evolution, and measurement. However, if one corner in the above diagram is
unknown or missing, then we need assumptions about the other two.}%
\label{triangle}%
\end{figure}
As shown diagramatically in Fig.\ \ \ref{triangle} we can divide any experiment
into preparation, evolution and measurement. If one of the elements of the
triad is unknown or missing then we need to previously characterize the other
two making assumptions in the process.

\subsection{Practical detector tomography}

In state tomography, one must perform a set of measurements $\{\pi
_{n}\}$ spanning the space of the density operator to be reconstructed. If the
state is defined on a $d$-dimensional Hilbert space, then it will be fully
specified by $d^{2}-1$ real parameters (respecting the constraint of unit
trace). To fully characterize a quantum detector, we need the data obtained
from measuring input states from a well-characterized source. To recover all
the POVM elements $\{\pi_{n}\}$ from the measured statistics $p_{\rho,n}$ the
probe states or input states must also be chosen to form a set $\{\rho_{j}\}$
that is tomographically complete: the span of the operators
$\{\rho_{j}\}$ -- which are not necessarily linearly independent -- must be
the entire space from which $\pi_{n}$ is taken. A spanning set forming an
operator basis will have at least $d^{2}(k-1)$ elements for a $k$ outcome
detector.
In principle this is sufficient to calculate the direct inversion of
Eq.\ (\ref{Trace}). However experimental detector tomography carries
additional requirements. Clearly the probe states should be previously
characterised, and large numbers of them should be easily and reliably
generated. In the case of optical detectors, coherent states are ideal
candidates since a laser can generate them directly and we can create a
tomographically complete set by transforming their amplitude through
attenuation (for example with a beam splitter) and a phase-shifter (an
optical path delay). Using input states $\{|\alpha\rangle\langle
\alpha|:\alpha\in\mathbbm{C}\}$ one can then reconstruct the $Q$-function of
the detector \cite{PRLSoto} which is simply proportional to the measured
statistics,
\begin{equation}
p_{\alpha,n}=\frac{1}{\pi^{2}}\langle\alpha|\pi_{n}|\alpha\rangle=\frac{1}%
{\pi}Q_{n}(\alpha).\label{probpure}%
\end{equation}
Since $Q_{n}(\alpha)$ of each POVM contains the same information as the
element $\pi_{n}$ itself, predictions of the detection probabilities for
arbitrary input states can then be calculated directly from the Q-function.

\subsection{Simple example: the perfect photocounter}

Consider as a simple example the case of a perfect photon number detector
This projective measurement is characterized by its POVM elements, $\left\{
\pi_{n}=|n\rangle\langle n|:{n=0,\dots,N}\right\}  $. In the simplest of
scenarios, using pure number states, $\left\{  \rho=|m\rangle\langle
m|:{m=0,\dots,N}\right\}  $, the characterization would be trivial, since the
statistics
\[
p_{\rho,n}=\delta_{n,m}%
\]
would immediately characterize our detector. From a more practical
perspective, pure number states are very hard to generate, especially
for high photon numbers. Using coherent states is therefore a more realistic
approach. Assuming a collection $\{|\alpha_{i}\rangle:i=1,\dots,D\}$ of perfect
coherent states our statistics would then become
\begin{align*}
p_{\alpha_{i},n}  & =\mathop{\rm Tr}{_{}}\left(  |\alpha_{i}\rangle
\langle\alpha_{i}| n\rangle\langle n|\right)  \\
& =e^{-{|\alpha_{i}|}^{2}/2}\frac{|\alpha_{i}|^{2n}}{n!}.
\end{align*}
Of course in an experiment we would not know the POVM elements in advance, and
we would need to express them with a generic expression such as
\[
\pi_{n}=\sum_{k,p}\theta_{k,p}^{(n)}|k\rangle\langle p|
\]
suffering only the constraint $0 \leq \pi_n \leq \mathbbm{1}$.
Fortunately, these operators can be simplified: interposing a phase
shifter in the coherent beam's path we can check if the statistics
are independent of the phase.  If they are we can
infer the phase independence of the
POVM. Our operators then become $\pi_{n}=\sum_{k}\theta_{k}^{(n)}%
|k\rangle\langle k|$ and the statistics can be expressed as
\begin{equation}
p_{\alpha_{i},n}=e^{-{|\alpha_{i}|}^{2}/2}\sum_{k=0}^{\infty}\frac{|\alpha
_{i}|^{2k}}{k!}\theta_{k}^{(n)}.\label{simple_prob}%
\end{equation}

\begin{figure}[ptbh]
\begin{center}
\includegraphics[width=9cm]{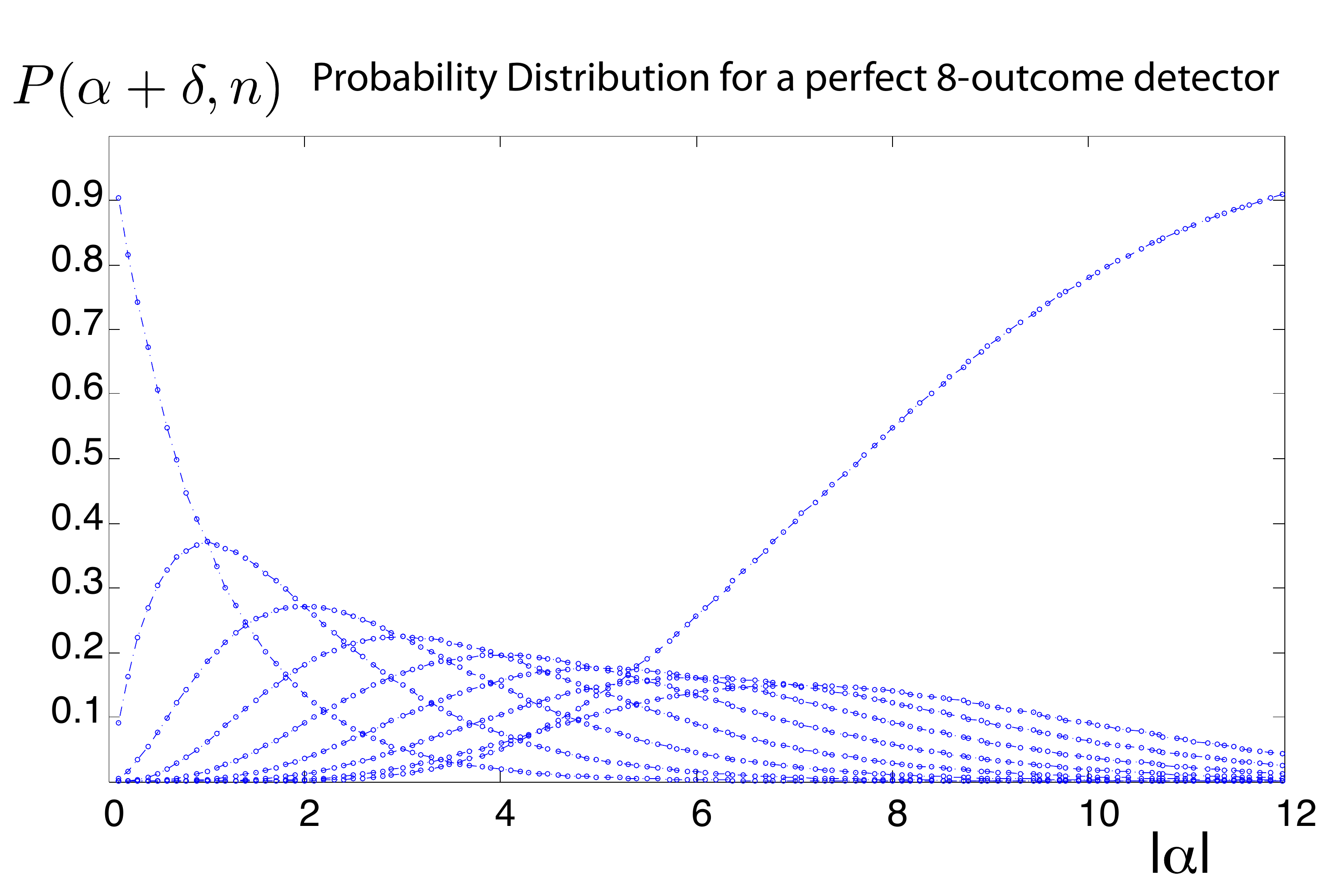}\\[0pt]
\end{center}
\caption{Outcome probability distributions for an $8$-outcome detector.
Each curve represents the probability of that outcome (zero clicks, one click, etc) happening
vs. the value of the intensity of the coherent state arriving at the detector.}%
\label{simplePD}%
\end{figure}
For a perfect photon number detector that can discriminate up to
eight photons, the outcome probability distributions would be the one
shown in Fig.\ \ref{simplePD}. 

We now consider how one would estimate the POVM elements from these
probability distributions, which form a set of simple simulated data.
Our goal is to invert Eq.\ (\ref{simple_prob}). To do so we can write a
matrix version of the equation. Given the set of coherent states $\{\alpha
_{1},\dots,\alpha_{D}\}$, and truncating the number states at a sufficiently
large $M$, we can write
\begin{equation}
{P}={F}{\Pi}.\label{matrix_equation}%
\end{equation}
Here, we have taken
\begin{align*}
p_{\alpha_{i},n} =  & \langle\alpha_{i}|\hat \pi_n |\alpha_{i}\rangle\\
& =\langle\alpha_{i}|\biggl(\sum_{k=0}^{M}\theta_{k}^{(n)}|k\rangle\langle
k|\biggr)|\alpha_{i}\rangle\\
& =e^{-|\alpha_{i}|^{2}}\sum_{k=0}^{M}\frac{|\alpha_{i}|^{2k}}{k!}\theta
_{k}^{(n)}\\
& =\sum_{k=0}^{M}F_{k}(\alpha_{i})\theta_{k}^{(n)}\\
& =\sum_{k=0}^{M}{F}_{i,k}{\Pi}_{k,n}.
\end{align*}
So the matrix $P$ has entries
\[
P_{i,n}= p_{\alpha_{i},n},
\]
$F$ entries ${F}_{i,k}=F_{k}(\alpha_{i})$, and $\Pi$ entries ${\Pi}%
_{k,n}=\theta_{k}^{(n)}$. For an $N$-outcome detector, this gives the matrix
dimensions of $P\in\mathbbm{C}^{D\times N}$, $F\in\mathbbm{C}^{D\times M}$,
and $\Pi\in\mathbbm{C}^{M\times N}$.

\begin{figure}[ptbh]
\begin{center}
\includegraphics[width=9cm]{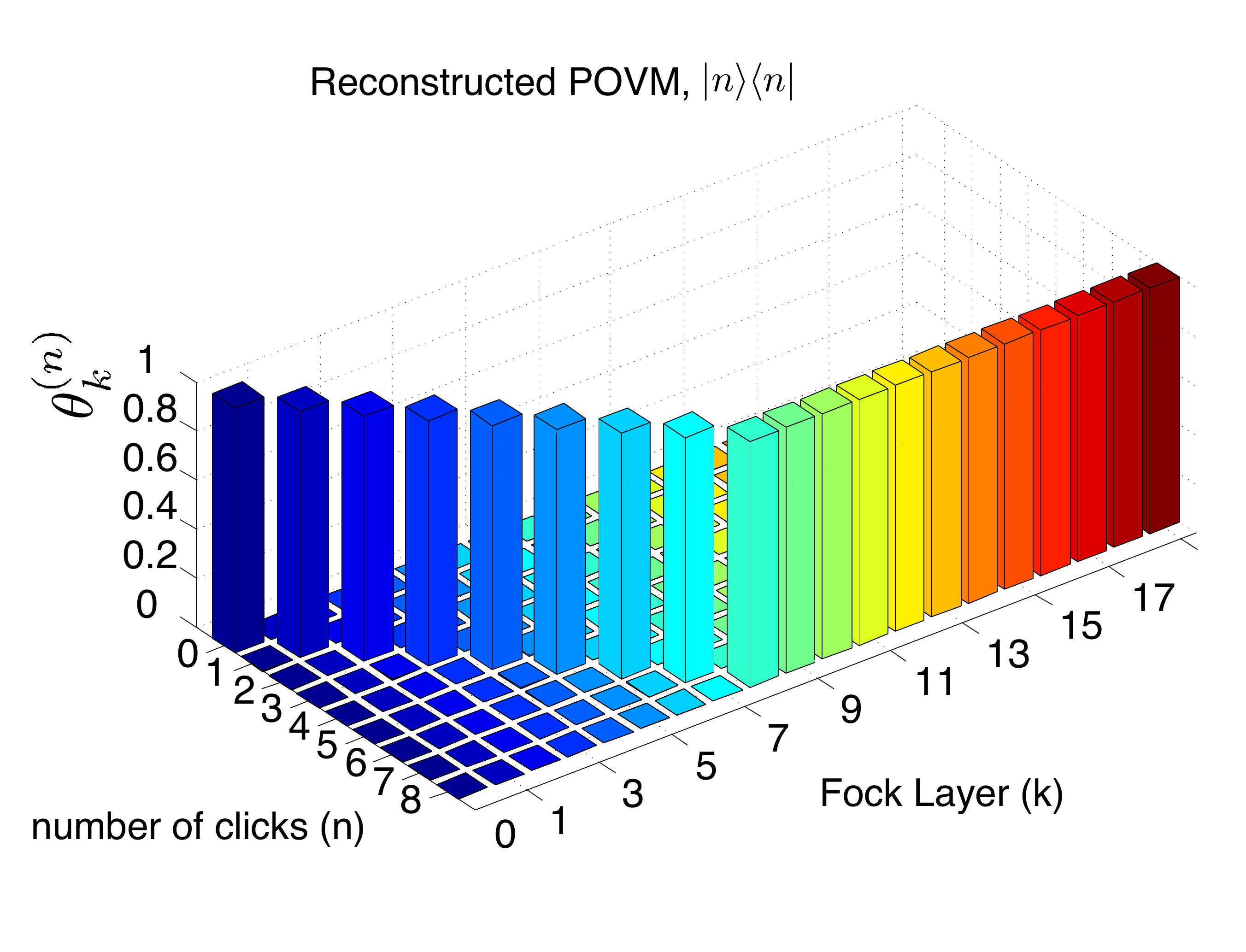}\\[0pt]
\end{center}
\caption{Reconstructed POVM of a theoretical photon number detector with $9$
bins. The POVM elements, $\pi_{n} = \sum_{k} \theta_{k}^{(n)} |k\rangle\langle
k|$, are the result of the optimization in Eq.\ (\protect\ref{simple_optim}), and show
a perfect result even though a $2$\% random noise was added to the data
(changing the value of the coherent state amplitudes). }%
\label{simplePOVM}%
\end{figure}

Now, to obtain $\Pi$, we can simply solve the convex optimization problem
\begin{align}
\label{simple_optim}\text{min} & \left\{  \|{P}-F \Pi\|_{2}\right\}
,\nonumber\\
\text{subject to }  &  \; \pi_{n} \geq0,\,\;\;\sum_{n=1}^{N} \pi_{n}
=\mathbbm{1}.
\end{align}
This is a convex problem, because the norm $\|.\|_{2}$, defined as
$\|A\|_{2}=(\sum_{i,j} |A_{i,j}|^{2})^{1/2}$ for matrix $A$ is a convex
function and the positivity constraint $\pi_{n}\geq0$ is semi-definite. The
result of a single such minimization is shown in Fig.\ \ref{simplePOVM}, where
we recover the expected POVM elements
\begin{equation}
\biggl\{ |0\rangle\langle0|, |1\rangle\langle1|, ,\dots, .,|7\rangle\langle7|,
\mathbbm{1}- \sum_{k=0,\dots, 8} |k\rangle\langle k| \biggr\}.
\end{equation}
It is remarkable that even introducing some statistical noise in the simulated
data the results are just as perfect. Indeed, if instead of using
$p_{\alpha_{i},n} = \mathop{\rm Tr}{_{}}\left( |\alpha_{i}\rangle\langle
\alpha_{i}| n\rangle\langle n|\right)  $ we use, $p_{\alpha_{i},n} =
\mathop{\rm Tr}{_{}}\left( |\alpha_{i}+\delta_{i}\rangle\langle\alpha
_{i}+\delta_{i}| n\rangle\langle n|\right)  $ where $\{\delta_{i}\}$
represent a $2$\% random noise, then the results are just as robust as without. This will
be discussed in more detail in later sections as it relates to the technical
noise of the laser. Let us then move on to more realistic detectors and see
the problems that loss and finite photon resolution introduce.

\section{Why detector tomography?}

In spite of its first sucessful applications, one could question the
need for detector tomography.  After all, detectors have been
calibrated in the past without tomographic techniques.  However, as
quantum technology makes striking advances, quantum detectors are
becoming more complex and, thus, susceptible to imperfections that are
not incorporated in the bottom-up approach of modelling. In contrast,
with tomography one can design detectors with a top-down aproach by
fully characterizing the final detector operation. For example,
photo-detection has seen the advent of single-carbon-nanotube
detectors \cite{nanotube-detector}, charge integration photon
detectors (CIPD) \cite{CIDP}, Visible Light Photon Counters (VLPC)
\cite{VLPC}, quantum dot arrays \cite{shields:3673}, superconducting
edge and picosecond sensors \cite{miller:791,super-pico} or time
multiplexing detectors based on commercial Si-APDs
\cite{loopy,loopy0}. Certainly understanding and modelling in full
detail the noise, loss and coherence characteristics of these
technologies is far from trivial. Detector tomography is an answer to
those challenges and, additionally, will allow us to benchmark
competing detectors.  Such characterized detectors also allow for the
preparation of non-Gaussian states in a certified manner, such as
photon subtracted states. Hence, such detectors are readily useable in
protocols such as entanglement distillation based on non-Gaussian
states \cite{Dis1,Dis, Dis2}, in schemes increasing entanglement by means of
photon subtraction \cite{In}, enhancing the teleportation fidelity
\cite{Tel1,Tel2}, or in other applications of non-Gaussian states
\cite{Loophole,Met}, specifically in the context of quantum metrology.

Let us discuss more precisely how detector tomography can provide
an advantage with respect to traditional calibration methods.

\subsection{The limits of calibration}

Standard calibration methods require a model of the detector.  The
parameters of this model are then estimated experimentally using
states and standard assumptions.  However, for specific applications, 
this models can become very complicated with a daunting number of parameters.  
For example, a detector as simple as a Yes/No avalanche photo diode (APD) becomes
very hard to model when all noise sources are studied \cite{migdall1,migdall2}.
Tomography sidesteps this by enabling the operational detector to be measured directly.
Another advantage can be to avoid errors or pitfalls from standard calibration.
An example is the use of Bell-state detectors which can appear to be working
while frequency correlations in the input states obscure the results \cite{Broken-Bell-State-Detector,Ian_royal_soc}.
To avoid such pitfalls detectors are often used in the rages where their behaviour is easily calibrated.
Detector tomography could extend the range of applicability of existing detectors and help design more complicated ones.
Let us now look at other advantages of full detector tomography.

\subsection{Quantitative entanglement verification}

Once a detector is fully characterized, it can be used to characterize
states in a certified fashion. In this respect, a detector is still useful if 
it is imperfect in the sense that its POVM elements are not just
unit rank projections: One can use them in an estimation
problem, or in a setting that unambiguously estimates properties of a
state. This is one of the key applications of detector tomography, in
that imperfect devices can be used to very reliably perform
estimation. A specifically important example is the direct detection
of entanglement: Once detectors are characterized, one can perform
measurements and then find lower bounds to entanglement measures,
based on these measurements. This is an idea presented in Ref.\
\cite{Auden} (see also Refs.\ \cite{Quant,GuehneShort,GuehneLong,Wunder-Plenio}).
Here we discuss the specific issues related to infinite-dimensional
systems and photon counters.

Assume that we have performed more than one type of measurement,
labeled by $k$, for which we have completely characterized the POVM elements
$\{\pi_{n}^{(k)}\}$ to great accuracy. Using two such devices
 one can make local measurements on each part of a bipartite
state. With the data from
\[
d_{n,m}^{(k,l)}=\mathop{\rm Tr}{_{}}\left(  (\pi_{n}^{(k)}\otimes\pi_{m}%
^{(l)})\rho\right)
\]
one can then ask for the minimal degree of entanglement consistent with it.
This approach does not make any assumptions on the preparation of states, and
works even for detectors with a low detection efficiency; This will
already be incorporated in the stated bound. The unambiguously minimal degree
of entanglement, say in terms of the negativity, is then given by the solution
of the optimization problem \cite{Auden}:
\begin{align*}
\text{min}\, &  E_{N}(\rho)=\Vert\rho^{\Gamma}\Vert_{1}-1,\\
\text{subject to } &  \mathop{\rm Tr}{_{}}\left(  (\pi_{n}^{(k)}\otimes\pi
_{m}^{(l)})\rho\right)  =d_{n,m}^{(k,l)},\\
&  \mathop{\rm Tr}{_{}}\left(  \rho\right)  =1,\,\rho\geq0.
\end{align*}
One easily finds lower bounds to the optimal solution to this problem by
considering
\[
P=\sum_{n,m,k,l}\alpha_{n,m}^{(k,l)}\pi_{n}^{(k)}\otimes(\pi_{m}^{(l)}%
)^{T}+\beta\mathbbm{1}
\]
such that $\Vert P\Vert\leq1$. Then a lower bound is readily given by
\cite{Quant}
\begin{align*}
\Vert\rho^{\Gamma}\Vert_{1}  & \geq\mathop{\rm Tr}{_{}}\left(  \rho^{\Gamma
}P\right)  =\mathop{\rm Tr}{_{}}\left(  \rho P^{\Gamma}\right)  \\
& =\sum_{n,m,k,l}\alpha_{n,m}^{(k,l)}d_{n,m}^{(k,l)}+\beta.
\end{align*}
The optimal lower bound, based on such an approach, is in turn the solution of
the convex optimization problem \cite{Convex} in $\alpha_{n,m}^{(k,l)}$ and
$\beta$ given by
\begin{align*}
\text{max}\, &  \sum_{n,m,k,l}\alpha_{n,m}^{(k,l)}d_{n,m}^{(k,l)}+\beta,\\
\text{subject to } &  -\mathbbm{1}\leq\sum_{n,m,k,l}\alpha_{n,m}^{(k,l)}%
\pi_{n}^{(k)}\otimes(\pi_{m}^{(l)})^{T}+\beta\mathbbm{1}\leq\mathbbm{1}.
\end{align*}
Some conditions need to be examined to make sure that the solution to
this formulation (dual optimal) coincides with the solution of the original
formulation (primal optimal).  For linear programs or for 
 semi-definite problems (SDP) as the one described these conditions are
easily satisfied \cite{Convex}.  As we will see this type of optimisation
method is also useful for detector tomography itself.

Here, if one just makes use of POVM elements with a finite support, as
one has in photon counting with an additional phase reference, then such
bounds will provide very strong lower bounds.  However, it will not
provide good bounds for unbounded operators such as in homodyning. Hence,
without having detectors of high efficiency, and without assumptions on the
preparation of the entangled state, one can use a characterized detector to
certify entanglement in quantitative terms.

\section{Modelling photodetectors}

As we have seen, the aim of detector tomography is to identify the
physical POVM closest to the experimental data with minimal
assumptions on the functioning of the detectors.  To compare this
method with a more traditional calibration method let us study a
photodetector modelling example.  We will do so for an avalanche
photodiode and a photon number resolving detector able to detect up to
$8$ photons \cite{loopy}.  In the next section we will compare these models to
the experimental results derived without any model.

\subsection{Optical photon number detectors}

An important detector in quantum optics is the single-photon counting module
based on a silicon avalanche photodiode (APD). It has two detection outcomes,
either registering an electronic pulse ($1$-click) or not ($0$-clicks). A
loss-free perfect version of it would implement the Kraus operators
\begin{equation}
\{ |0\rangle\langle0|, \mathbbm{1}-|0\rangle\langle0|\},
\end{equation}
distinguishing between the presence or absence of photons. However some
photons are absorbed without triggering a pulse. This loss can be modelled
placing a BS in front of the perfect detector \cite{Ulf-loss}. The POVMs
describing a detector with a BS of transmittivity $\eta$ can then be written
as,
\begin{align}
\text{\textsc{no click}}  & : \pi_{0}=\sum_{n=0}^{\infty}\left(
1-\eta\right)  ^{n}\left\vert n\right\rangle \left\langle n\right\vert ,\\
\text{\textsc{click}}  & :\pi_{1}=\mathbbm{1}-\sum_{n=0}^{\infty}\left(
1-\eta\right)  ^{n}\left\vert n\right\rangle \left\langle n\right\vert .
\end{align}
disregarding after-pulsing or dark counts \cite{APD-Hendrik}. Having only two
outcomes, this detector cannot distinguish the number of photons present.

\label{TMD-model} A more advanced detector called time multiplexing
detector (TMD) does have certain photon-number resolution. It splits
the incoming pulse into many temporal bins, making unlikely the
presence of more than one photon per bin. All the time bins are then
detected with two APDs. Summing the number of $1$-click outcomes from
all the bins one can then estimate the probability of having detected 
a number of incoming photons. This detector is not commercially
available but one has been constructed by the Ultrafast Group in Oxford
\cite{loopy}. It has eight bins in total (four time bins in each of
two output fibres) and thus nine outcomes -- from zero to eight
clicks.\\\\

\begin{figure}[ptbh]
\begin{center}
\includegraphics[width=6cm]{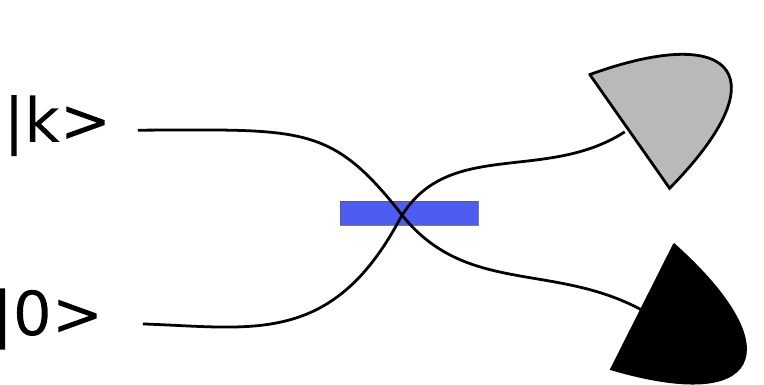}\\[0pt]
\end{center}
\caption{ Diagram of a simplified $2$-bin multiplexing detector. }%
\label{2bin-loopy}%
\end{figure}

The theoretical description of this detector is a bit more involved since
there is what we call the ``binning problem''. Indeed, in addition to loss
there is a certain probability that all photons will end up in a single time
bin, or more generally that $k$ incoming photons will result in less than $k$
clicks. To account for the details describing these probabilities we use a
recursive relation \cite{martin-private,loopy-achilles}. Our goal is to
describe the following probability distribution:

$p^{N}(j/k)$: Probability of having $j$-clicks given that there were $k$
incident photons and that the detector has $N$-bins (or modes). Let us start
with the simplest possible TMD which would consist of an input port, a beam
splitter (with reflectivity and transmittivity $R$ and $T$) and two YES/NO
detectors. This detector is shown in Fig.\ \ref{2bin-loopy} and has two bins.

For this simple example, the probability distribution we are after is
$p^{2}(j/k)$. We will show how to calculate $p^{2}(j/k)$, $p^{4}(j/k)$ and
then how to go from $p^{N}(j/k)$ to $p^{2N}(j/k)$. For the two bin case from
Fig.\ \ref{2bin-loopy} consider a BS with transmittivity $T$ and reflectivity
$R$. In that case:

\begin{itemize}
\item $p^{2}(j,0) = \delta_{j,0}$ (if no photons are present we will only
register zero clicks).

\item $p^{2}(1,k) = T^{k}+R^{k}$ (with probability $T^{k}$, $k$ photons end up
in the lower bin and the same holds for the upper bin with $R^{k}$. The
probability of a single click is the sum of these independent probabilities).

\item $p^{2}(2,k) = 1 - T^{k}+R^{k}$ (if $k \neq0$ then only two events may
happen: one click or two. This complementary event has therefore $P = 1 -$
(Probability of $1$ click).)
\end{itemize}

\begin{figure}[ptbh]
\begin{center}
\includegraphics[width=7cm]{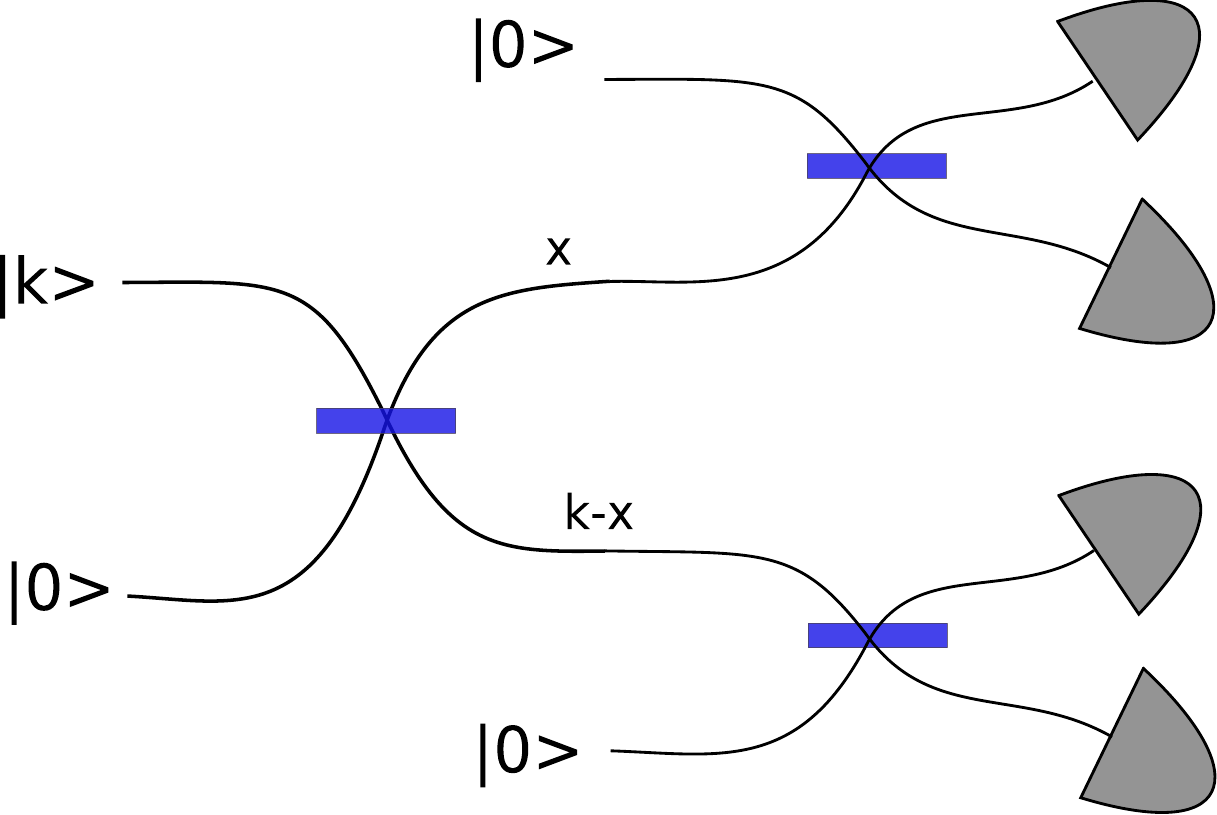}\\[0pt]
\end{center}
\caption{Diagram of a simplified $4$ bin multiplexing detector. The first beam
splitter distributes $k$ photons according to a binomial distribution between
the two $2$-bin loopies of the second stage. }%
\label{4bin-loopy}%
\end{figure}

In the case of a $4$-bin detector shown in Fig.\ \ref{4bin-loopy}, $k$
incoming photons are distributed to two $2$-bin detectors according to a
binomial distribution.

Now let us evaluate the probability for the upper $2$-bin detector to register
$s$ counts if $x$ photons entered while registering $m$ clicks in the lower
2-bin detector if $k-x$ entered the lower port. This should be $p^{2}(s,x)
p^{2}(m,k-x)$ weighted by the probability that $x$ photons enter the upper
branch and $k-x$ the lower one, which is
\[
\binom{k}{x} T^{k-x}R^{x}.
\]
Now the probability that $j$ counts are found overall is found summing the
weighted probability over all possible ways that the detectors can find $j$
counts (i.e., $m+s=j$) and summing over all possible ways of distributing $k$
photons:
\[
p^{4}(j/k) = \sum_{x=0}^{k} \sum_{m+s=j} \binom{k}{x} T^{k-x} R^{x} p^{2}(s,x)
p^{2}(m,k-x).
\]

\begin{figure}[ptbh]
\begin{center}
\includegraphics[width=7cm]{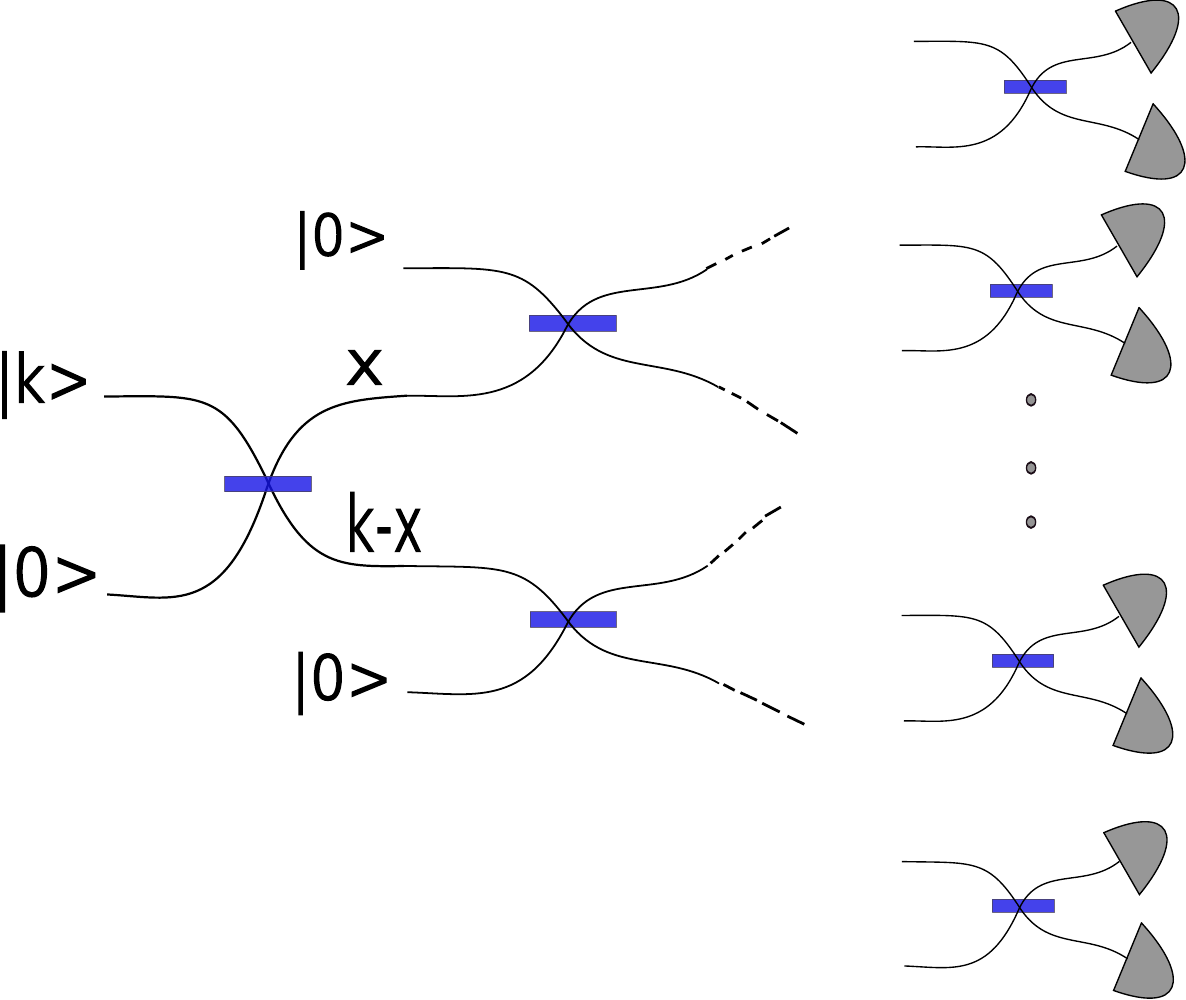}\\[0pt]
\end{center}
\caption{Diagram of a $2N$-bin multiplexing detector. The first beam splitter
distributes $k$ photons according to a binomial distribution between the two
next $N$-bin stages. }%
\label{2nbin-loopy}%
\end{figure}

We can extend the same argument to $2N$. Imagine we know $p^{N}(j/k)$. Now,
for that detector to become a $2N$-bin detector all we need is to couple two
of them to a beam splitter which will distribute the $k$ photons as described
above and as shown in Fig.\ \ref{2nbin-loopy}. In that same fashion we can
then define the recursive relation

\begin{equation}
p^{2N}(j/k)=\sum_{x=0}^{k}\sum_{m+s=j}\binom{k}{x}T^{k-x}R^{x}p^{N}%
(s,x)p^{N}(m,k-x).\label{recursive-loopy}%
\end{equation}

This of course can be generalized to accommodate a different BS
reflectivity at each node (adding an index to $T$ and $R$ to account
for its position). Based on this recursion and once we determine all $T$ and
$R$, we can write a simple and efficient program to generate the corresponding
theoretical POVMs. For example, a $6$-outcome detector would have a POVM which
can be captured in the following matrix:

{ \footnotesize
\[%
\begin{array}
[c]{lcccccc}
& \pi_{0} & \pi_{1} & \pi_{2} & \pi_{3} & \pi_{4} & \pi_{5}\\
|0\rangle\langle0| & 1 & 0 & 0 & 0 & 0 & 0\\
\noalign{\medskip}|1\rangle\langle1| & 0 & 1 & 0 & 0 & 0 & 0\\
\noalign{\medskip}|2\rangle\langle2| & 0 & 0.128 & 0.871 & 0 & 0 & 0\\
\noalign{\medskip}|3\rangle\langle3| & 0 & 0.0168 & 0.334 & 0.648 & 0 & 0\\
\noalign{\medskip}|4\rangle\langle4| & 0 & 0.00226 & 0.100 & 0.495 & 0.400 &
0\\
\noalign{\medskip}|5\rangle\langle5| & 0 & 0.000309 & 0.0283 & 0.265 & 0.508 &
0.197\\
\noalign{\medskip}|6\rangle\langle6| & 0 & 0.0000428 & 0.00772 & 0.123 &
0.421 & 0.447\\
\noalign{\medskip}|7\rangle\langle7| & 0 & 0.00000601 & 0.00208 & 0.0536 &
0.291 & 0.653\\
\noalign{\medskip}|8\rangle\langle8| & 0 & 0.000000852 & 0.000565 & 0.0224 &
0.181 & 0.794
\end{array}
\]
} 

where $B_{k,j}=p^{5}(j/k)$, $\;j=0,\dots,5$ and $k=0,\dots,8$. For example,
the $5$-click event has a POVM element,
\[
\pi_{5}\simeq0.2|5\rangle\langle5|+0.4|6\rangle\langle6|+0.6|7\rangle
\langle7|+0.8|8\rangle\langle8|,
\]
etc. More generally, the measured statistics are related to the incoming
photons by
\[
p_{j}=\sum_{k}p^{N}(j/k) \; q_{k},
\]
where $p_{j}$ is the probability of detecting $j$ counts and $q_{k}$ the
probability that $k$ photons arrived to the TMD \cite{loopy0}. The Matrix
$\mathbf{B}$ introduced above then relates probabilities and density matrices
through: $\mathbf{p}=\mathbf{B}\cdot\mathbf{\rho}$.


\subsection{Detector loss}

\begin{figure}[ptbh]
\begin{center}
\includegraphics[width=8cm]{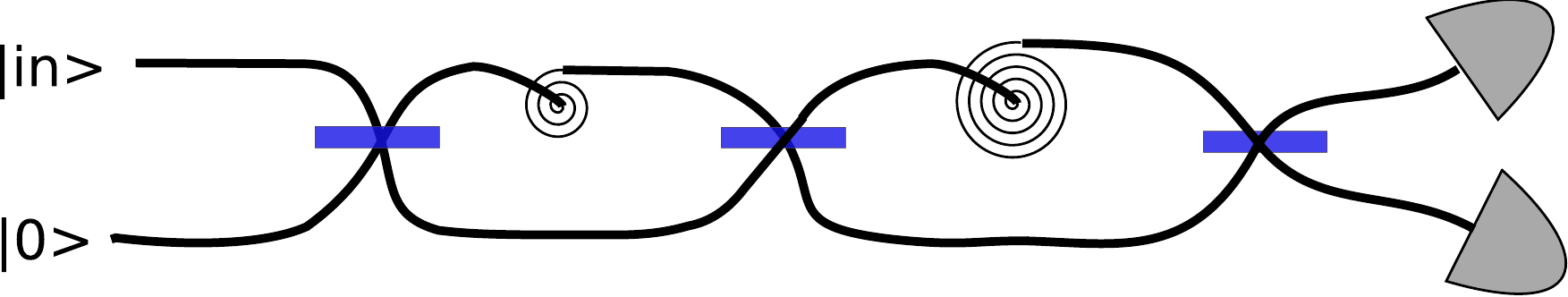}\\[0pt]
\end{center}
\caption{ Diagram of a simplified 8 bin time multiplexing detector (TMD). The
spirals represent a delay in the optical fibre. }%
\label{loopy-diagram}%
\end{figure}

TMD detectors have various sources of loss, meaning that photons are absorbed
before triggering a detection event. The major sources of loss are the
coupling to the fibres, the absorption and scattering in the delay fibres and
the non-unit efficiency of the detectors \cite{achilles-2006-97}. A full
description of the effect of losses is certainly complex, since loss occurs at
many stages of the detector. One can model loss simply as a beamsplitter
diverting photons out of an input state. However, one would have to include a
BS before the detector (fibre coupling), a BS at each stage of fibre and a BS
in front of each APD, altering Eq.\ (\ref{recursive-loopy}) accordingly.
Instead we give an effective description modeling loss with a single BS in
front of the detector. The matrix capturing the losses has entries that are
given by
\[
L_{k^{\prime},k}=\binom{k}{k^{\prime}}\eta^{k^{\prime}}(1-\eta)^{k-k^{\prime}%
},
\]
being the binomial distribution accounting for loss since $L_{k^{\prime},k}=0$
for all $k<k^{\prime}$. Now combining both descriptions, we can decouple the
loss from the binning, putting a BS coupled to the environment before the
$N$-bin TMD resulting in
\[
p_{j}=\sum_{k,k^{\prime}}p^{N}(j/k^{\prime})L_{k^{\prime},k}\; q_{k}.
\]
This relationship expresses how the incoming photons experience loss and then
are distributed among the available modes. The model of the TMD described up
to this point would be the one needed without detector tomography. One could
for example try to measure independently the transmittivities of the inner BS,
reconstruct the convolution (or binning) matrix $\mathbf{B}$ and try to
estimate the overall loss to include it in the matrix $L$. This would help us
build a model of the TMD sketched in Fig.\ \ref{loopy-diagram}. By contrast,
using detector tomography, we do not need to know anything about bins, beam
splitters, inner detectors or specific loss mechanisms. Moreover,
anything left out of our detector model (e.g. dark counts, afterpulsing, etc.)
would be included in a tomographic characterization.

\section{Experimental reconstruction}

We now turn to the description of the experimental realisation. As mentioned
earlier, coherent states are ideal probe states with which to characterize
optical detectors. This holds true for any optical detector, including
polarization detectors, frequency detectors (e.g. spectrometers), and even
detectors that discriminate inherently quantum states (e.g. a photonic
Bell-state detector). In most of these cases, we are only interested in the
detector's action at a particular input photon number $n$ (usually $n=1$).
Still, we can reconstruct the detector POVM in the full photon number basis
and then restrict our attention to a particular subspace. It is interesting
that an optical state which can be fully described in a classical theory of
electromagnetism can be used to characterize uniquely quantum detectors. To
characterize a completely unknown detector (i.e. a black box) one would need
to vary all the available parameters of the probe set of coherent states:
spatial-temporal mode, polarization, phase, and amplitude (ensuring a
tomographically complete set of states is constructed). However, given that
frequency, time, position, momentum, and amplitude have infinite range, this
is impossible in practice. Consequently, one is required to make realistic
assumptions about the range of operation and sensitivity of any unknown
detector. One might additionally neglect some of these optical parameters if
one is only interested in a particular aspect of the full characterization.

The subjects of our detector tomography, the APD and TMD, possess the unique
features of single-photon sensitivity and photon-number resolution,
respectively. To characterize these features we vary only the amplitude and
phase of probe coherent states, while fixing the spatial-temporal mode. In
particular, the input wavepackets have a time extent shorter than the time
window of the electronics, and the center wavelength is within range of the
detectors. Light is coupled to both types of detector through single-mode
optical fiber, eliminating the possibility of any variation in the position or
momentum mode of the light. Critically, for the detectors to perform as
intended and in order to ensure the detectors are memoryless, the wavepackets
must be preceded and followed by time intervals in which there is no input
light. The APD is known to have a deadtime of roughly $50$ ns; a detection
that occurs before the input wavepacket arrives at the detector will make the
detector insensitive to the probe coherent state. The TMD splits the incoming
wavepacket into time bins spread over $500$ ns. The inverse of these two
timescales then sets an upper limit on the rate at which we can send probe
states to the detectors. We further limit the rate to ensure the detectors do
not heat up, which would change their properties over time. These time variant
features of the detectors could be illuminated with detector tomography but
this would be unnecessarily complicated, as they are quickly evident without
use of the full tomography procedure. \newline\newline When operated with a
gain high above their lasing threshold, lasers produce light well approximated by a coherent
state. Coherent states (and statistical mixtures thereof) are unique amongst pure
optical states in that, at a beamsplitter, the transmitted and reflected
states are unentangled. Consequently, through attenuation we can vary the
amplitude of a coherent state without changing its nature. In homodyne state
tomography, one must use a balanced detection technique to negate technical
noise in the laser. In contrast, by attenuating the laser light to the
single-photon level in our detector tomography scheme, the resulting coherent
state possesses an inherent amplitude uncertainty that renders the technical
noise insignificant.

\begin{figure}[ptbh]
\begin{center}
\includegraphics[width=8.5cm]{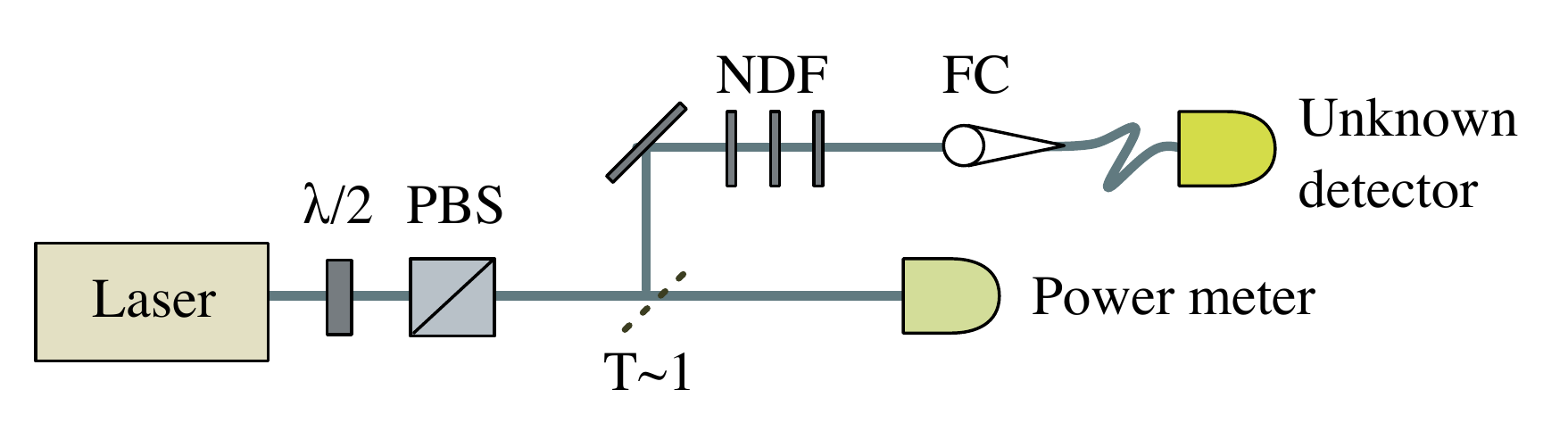}\\[0pt]
\end{center}
\caption{ Experimental setup:
The amplitude of the probe coherent state is attenuated with a half-waveplate ($\lambda/2$) and a Glan-Thompson polarizer (P).   The light is then further attenuated by Frequency independent filters or Neutral Density Filters (NDF) and coupled into a fibre (FC)  
 }%
\label{expsetup}%
\end{figure}

We use a cavity dumper (APE Cavity Dumper Kit) to reduce the repetition rate
of a pulsed mode-locked Ti:sapph laser to $R$. Long term drift of the laser
power over 1 million pulses was $<0.5\%$. Our laser randomly varies in energy
between subsequent pulses with a standard deviation of $1.88\%\pm0.02\%$ of
$\left\vert \alpha\right\vert ^{2}$. We vary the amplitude $\alpha$ of our
probe coherent states by rotating their polarization with a half-waveplate
($\lambda/2$) in front of a Glan-Thompson polarizer (P) as shown in
Fig.\ \ref{expsetup}. We attenuate the coherent states by reflecting them from
a beamsplitter (BS) ($T$=95\%) and three neutral density filters (NDF) (i.e.
spectrally flat attenuators). Note that $R/T$ for the BS was measured
with a relative deviation smaller than $1\%$.
Along with some loss upon coupling into a
single-mode fiber, we collect these attenuations together in an overall
attenuation factor $\gamma$. We test for any variation in $\gamma$ as a
function of $\alpha$ (e.g. which might be caused by rotating the waveplate if
it had a wedge) and find that the variation is less than $1\%$.

There are, as of yet, no direct techniques to calibrate the power of light at
the single-photon level \cite{Worsley09}. In fact, there are no laboratory
methods to make an absolute calibration of power at any intensity. Instead, a
chain of photo-detectors are calibrated relative to each other. At the
beginning of the chain is an absolute calibration system held at national
standards institutes. In the case of our power meter (Coherent FieldMaxII-TO),
the National Institute for Standards and Technology (NIST) uses a cryogenic
bolometer as its absolute standard. This chain results in a 5\% systematic
uncertainty in our measurements of the laser power $P$ (averaged over 0.2s),
measured at the transmitted port of the beamsplitter. The magnitude of
$\alpha$ for the probe state was found via $\left\vert \alpha\right\vert
^{2}=$ $\gamma P\lambda/(Rhc)$. For each value of $\alpha$ we recorded the
number of times each detection outcome occurred in $J$ trials (i.e. laser
pulses), which provides an estimate of $p_{\alpha,n}$. The phase of $\alpha$
was allowed to drift freely, during which no change in the $p_{\alpha,n}$ was
observed. Consequently, we did not actively vary the phase of our probe states.

The 5\% uncertainty in $P$ is the dominant error in our experiment. For a
detector with over 95\% efficiency, this error could result in estimates of
$\left\{  p_{\alpha,n}\right\}  $ that are incompatible with any physical
detector. For example, this could result in more detector clicks on average
than there were photons in the probe state on average. Gain in the detector
could achieve this, but at the same time would necessarily introduce noise
that would change the distribution of $p_{\alpha,n}$. For detectors with lower
efficiency, the systematic error in $P$ will simply add or subtract from
efficiency of the detector that results from the tomographic characterization.

We choose a center wavelength $\lambda$ and a FWHM bandwidth of $\Delta
\lambda$ that are appropriate for each detector. In the case of the APD
detector (a Perkin Elmer SPCM-AQR-13-FC) we set $\lambda=780\pm1$ nm,
$\Delta\lambda=20$ nm, and chose the appropriate rate $R=1.4975\pm0.0005$ kHz,
$J=1472967$, and $\gamma=(5.66\pm0.08)\times10^{-9}$. For the TMD detector we
set $\lambda=789\pm1$ nm, $\Delta\lambda=26$ nm, $R=76.169\pm0.001$ kHz,
$J=38084$, and $\gamma=(8.51\pm0.11)\times10^{-9}$.

Since $\left\vert \alpha\right\vert ^{2}$ has an infinite range, one must set
an upper limit on the magnitude of $\alpha$ used in our set of probe states. A
natural place for this is at the $\alpha$ at which the detector behavior
saturates, i.e. $p_{\alpha,n}$ stays constant as a function of $\alpha$. Since
this occurs asymptotically, a somewhat arbitrary degree of constancy must be
chosen; we set $\left\vert \alpha\right\vert ^{2}=120$ as our upper limit. We
expected the saturation behavior of the TMD would be $p_{\alpha,n}%
\approx100\%,$whereas we found that it was $p_{\alpha,8}=93.3\%$ and
$p_{\alpha,7}=6.7\%.$ Already, the measured statistics (which are proportional
to $Q_{n}(\alpha)$) give a clear signature that our theoretical model of the
detector is too simplistic. Moreover, as we increase the $\alpha$ beyond our
upper limit $p_{\alpha,8}$ begins to decrease. The cause for this behavior is
a break down of our memoryless detector assumption and highlights how crucial
it is to create probe states in the desired spatial-temporal mode. Along with
the probe laser pulse, the cavity dumper also out-couples a small fraction of
the preceding and following pulses in the Ti:Sapph cavity. These are separated
in time from our probe pulse by $13$ ns and each contain only $0.17\%$ the
energy of the probe pulse. Consequently, these extra pulses will have an
insignificant effect on $p_{\alpha,n}$ for most of the range of $\alpha$.
However, at $\left\vert \alpha\right\vert ^{2}=120$, the preceding pulse will
be a coherent state with $\left\vert \alpha\right\vert ^{2}=0.2$.
Consequently, roughly $20\%$ of the time bins in the TMD will be preceded by a
photon. If the detector has, for instance, a $50\%$ efficiency then $10\%$ of
the time bins will be preceded by a detection event that will not be counted
as click by the electronics (due to their time window). More importantly,
those bins will subsequently be unavailable to detect photons in our probe
pulse, due to the $50$ ns deadtime of the APDs inside the TMD. Thus, as we
observe, roughly $10\%$ of the bins will not result in a click. This behavior
was extraneous to the normal operation of the detector and so we limit
$\left\vert \alpha\right\vert ^{2}$ to $30$ in the tomographic reconstruction.
However, it exemplifies the usefulness of even the basic detector tomography
procedure, which results in approximate $Q_{n}(\alpha)$, for rough evaluation
of the detector action. Some of these hypothesis or details could be further
explored if we knew well some detectors or some states. For example the
response of BS and neutral density filters to single photons could be explored
(granted good single photons and reliable single photon detectors). The time
independence of the POVMs could also be studied with well known states.
However, we will see that the excellent fit of the data to the Q-function and
of the reconstruction to the model suggests a sufficiently good understanding.

\subsection{Results}

We now turn to the tomographic reconstruction. To characterize
our detector we have measured the outcome probability distributions
resulting from sending a tomographically complete set of input states (or probe states).
The use of pure coherent states as probe states
implies that the probability distribution is proportional to the Q-function,
as seen in Eq.\ (\ref{probpure}). In principle the knowledge of the Q-function
is then sufficient to predict the measurement probabilities for any incoming
state since $Q(\alpha,\alpha*) = \langle \alpha| \rho |\alpha\rangle$ determine $\rho$ completely.

\begin{figure}[ptbh]
\begin{center}
\includegraphics[width=8cm]{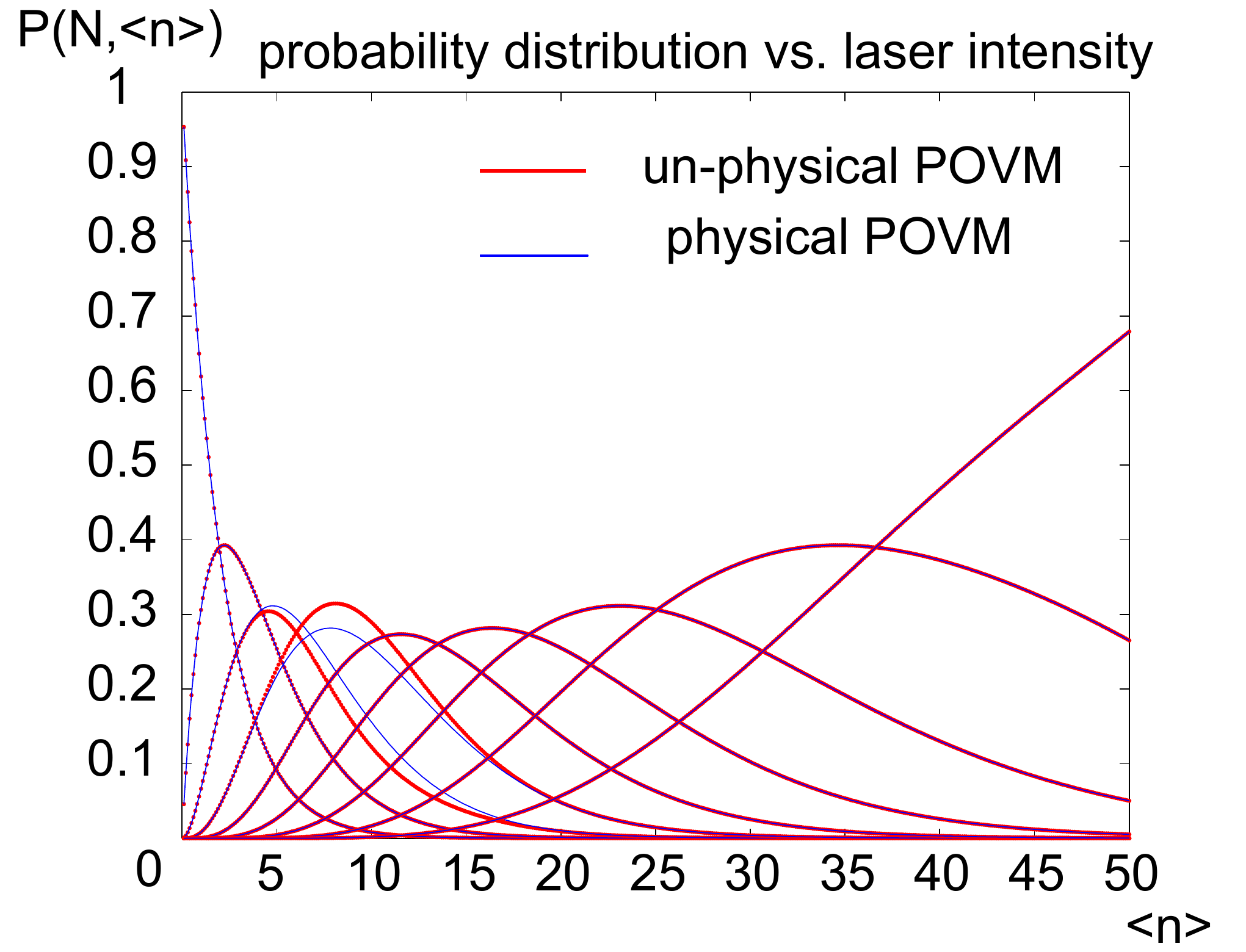}
\end{center}
\caption{This plot presents the probability distributions corresponding to two
detectors. One with negative \textquotedblleft POVM\textquotedblright%
\ elements and anotherone with positive ones. Both POVMs observe $\sum{\pi
_{n}}=\mathbbm{1}$. }%
\label{negativePOVM}%
\end{figure}

However, a more useful and natural representation for
photodetectors is the POVM expanded in the Fock basis. Another argument to
find the POVM elements is that due to statistical noise, the Q-function could
correspond to a non-physical POVM. Indeed, if we simply make a fit to the
noisy measured probability distribution, this fit could correspond to negative
\textquotedblleft POVMs\textquotedblright. As an example consider
Fig.\ \ref{negativePOVM} where two Q-functions are displayed together: 
Even though they are very similar to each other,
one corresponds to a detector with a non-positive \textquotedblleft
POVM\textquotedblright\ element (and thus negative probabilities) and one
corresponds to a physical one.  
Our goal is therefore to reconstruct the
POVM operators which most closely match the data and still observe 
Quantum Mechanics (and thus are positive).
. \newline\newline
Since we adopt a \textquotedblleft black box\textquotedblright\ approach we
need not assume any of the properties described in the previous theoretical
models. Only the accessible parts of the \textquotedblleft black
box\textquotedblright\, i.e., number of
outcomes and control (or lack of control) of the phase
will determine the description of our detector.
The lack of a phase reference simplifies the
experimental procedure, allowing us to solely control the magnitude of
$\alpha$ (as has been done for tomography of a single photon \cite{Lvovsky}).
A detector with no observed phase dependence will be described by POVM
elements diagonal in the number basis,%
\[
\pi_{n}=\sum_{k=0}^{\infty}\theta_{k}^{(n)}|k\rangle\langle k|,
\]
simplifying hence the reconstruction of $\pi_{n}$.


\begin{figure*}[ptbh]
\begin{center}
\includegraphics[width=12cm]{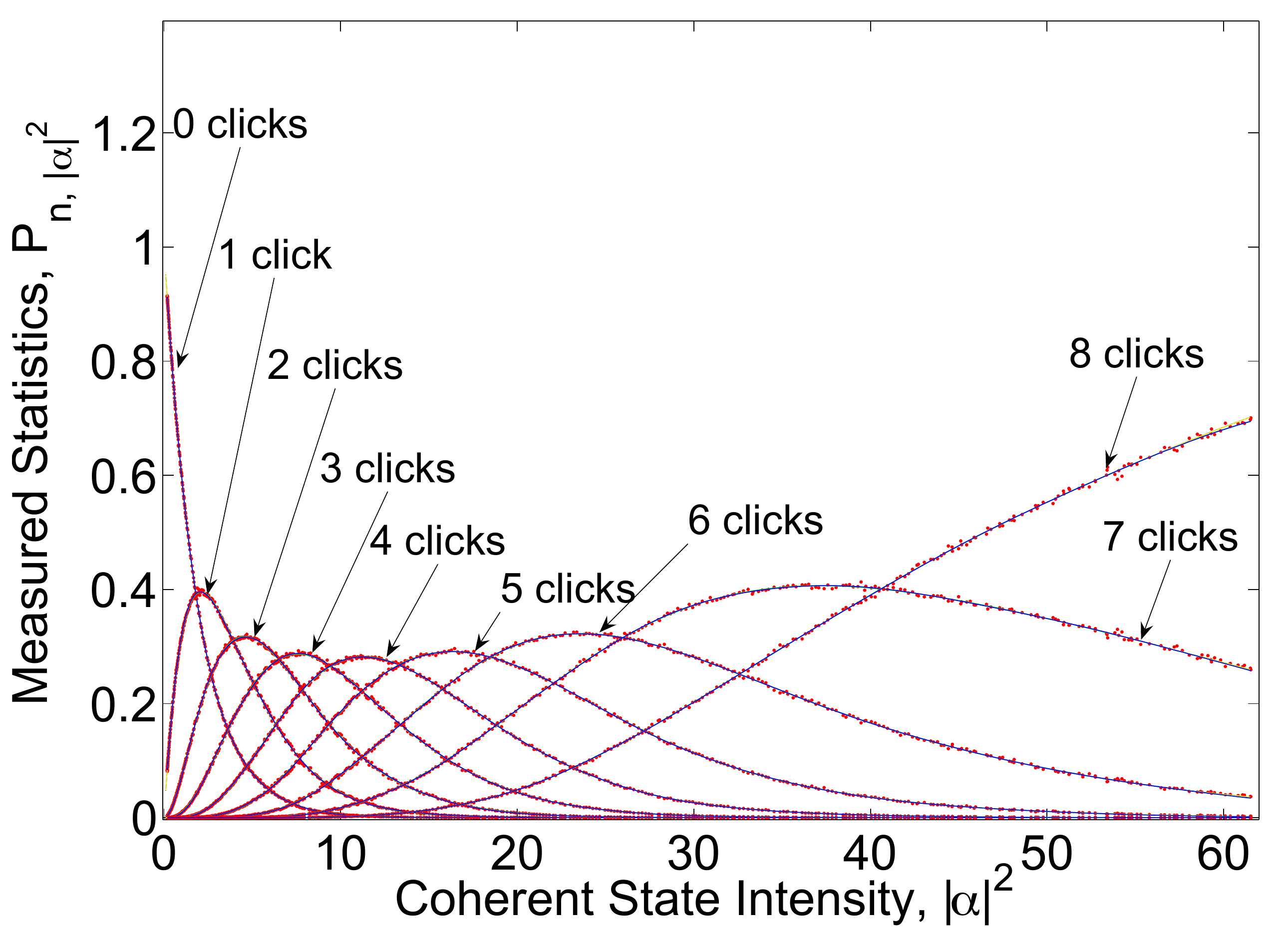}
\end{center}
\caption{ The measured probabilities
for different ``number-of-clicks'' are shown (red dots) as a function of
$|\alpha|^{2}=\langle n\rangle$. The plot shows the statistics for the time multiplexed
detector (TMD) with $9$ time-bins. The statistical error vertically is too small to be seen and the jitter of
$|\alpha|^{2}$ was estimated to be $2$\% of its value. An additional $5$\%
systematic error in the calibration of the power meter is present but can be
absorbed as loss. From the reconstructed POVM elements $\left\{  \pi
_{n}\right\}  $ we generate the corresponding probability distributions
$\text{tr}(\rho_{\alpha}^{\mathrm{(in)}}\pi_{n})$ (blue curves). These are
generated for pure $|\alpha\rangle\langle\alpha|$ or mixed $\rho_{\alpha
}^{\mathrm{(in)}}$ and for $\pi_{n}$ reconstructed with the filter function or
without it. For all these options, the probability distributions (blue lines)
are so similar that they are indistinguishable on this scale.}%
\label{qfun_TMD}%
\end{figure*}

Again we can express the relationship between the statistics and our diagonal
$\pi_{n}$ as,
\[
{P}=F \Pi,
\]
if we measure $D$ different values of $\alpha,\alpha_{1},\dots,\alpha_{D}$,
and truncate the number states at a sufficiently large $M$. For an $N$-outcome
detector, the matrices will have dimensions $P\in\mathbbm{C}^{D \times N}$,
$F\in\mathbbm{C}^{D \times M}$, and $\Pi\in\mathbbm{C}^{M \times N}$. In
addition
\[
F_{i,k} =\frac{ |\alpha_{i}|^{2k} \exp{(-|\alpha_{i}|^{2})} }{k!}%
\]
can easily be rewritten when the input state is a mixed state. This was done
indeed to account for the laser's technical noise (as we will see in the next
section) but gave similar results. For such a detector, the physical POVM
consistent with the data can be estimated through the following optimisation
problem:
\begin{align}
\label{optim} &  \text{min}\left\{  \|{P}-F \Pi\|_{2}+g(\Pi)\right\}
,\nonumber\\
&  \text{subject to } \;\; \pi_{n} \geq0,\,\;\;\sum_{n=1}^{N} \pi_{n}
=\mathbbm{1},
\end{align}
where the $2$-norm ensures it is a convex quadratic problem. Note that we also
allow for convex quadratic filter functions $g$ which will be discussed in some detail later.  These $g$ are related to the conditioning
of the problem and must not depend on the type of detector. For example, no
symmetry or knowledge of the typical POVM structures in photo-detection can be
assumed. If any, only general regularization functions that work for any
POVM should be chosen. Now, for suitable filter functions (i.e. cuadratic) the whole problem is a convex quadratic optimisation problem,
and hence also a semi-definite problem (SDP) which can be efficiently solved
numerically \cite{Convex}. Moreover, in this case, there exists a dual
optimisation problem whose solution coincides with the original problem. Thus,
the dual problem provides a certificate of optimality since it provides a
lower bound to the primal problem.

\begin{figure}[ptbh]
\begin{center}
\includegraphics[width=8.5cm]{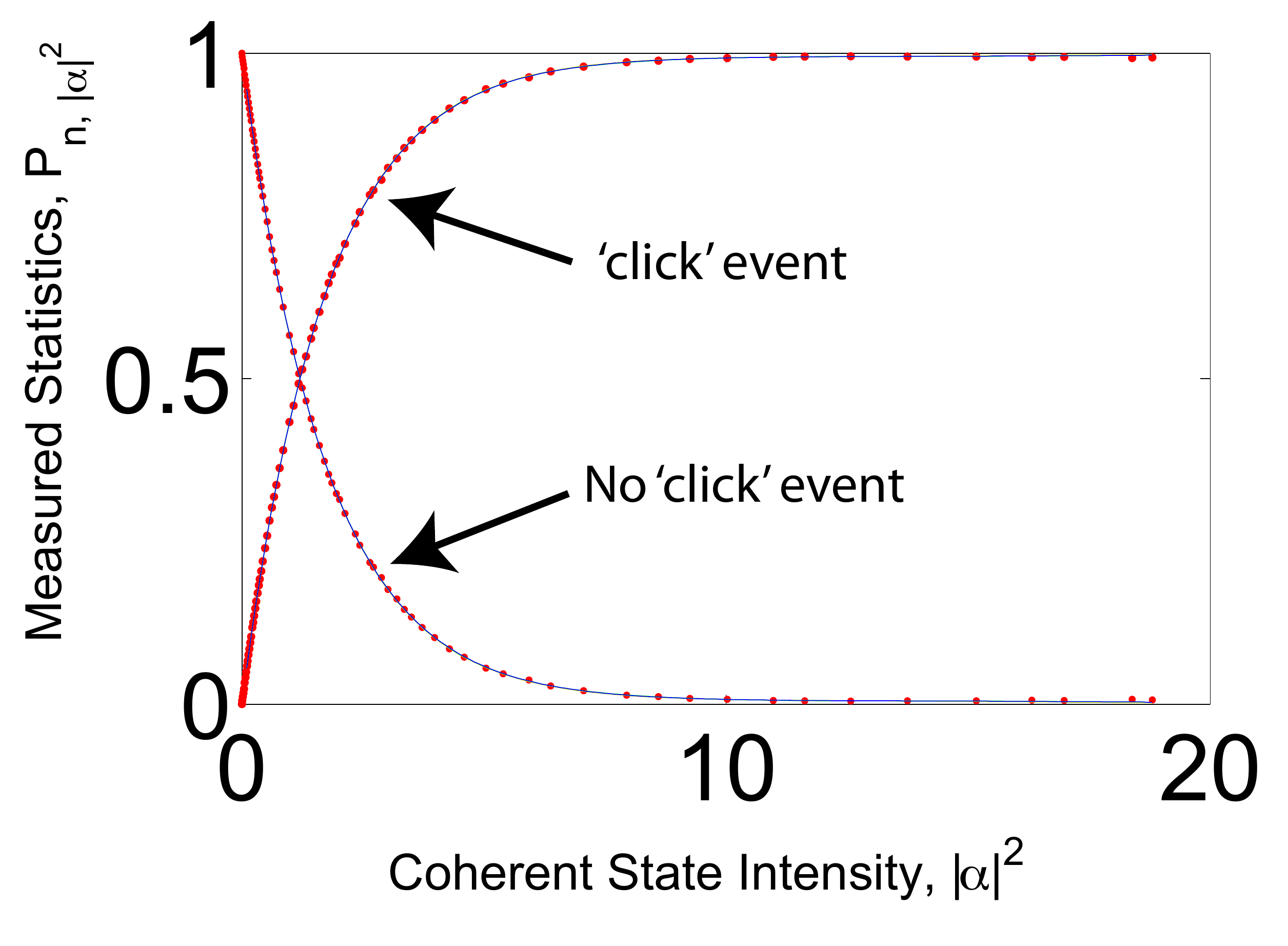}
\end{center}
\caption{The measured probabilities for the ``click'' and ``no-click''
envents in the Avalanche Photodiode (APD) are shown as a function of
$|\alpha|^{2}=\langle n\rangle$.  The statistical error vertically is too small 
to be seen and the jitter of $|\alpha|^{2}$ was estimated to be $2$\% of its value. 
An additional $5$\% systematic error in the calibration of the power meter is present but can be
absorbed as loss. From the reconstructed POVM elements $\left\{  \pi
_{n}\right\}  $ we generate the corresponding probability distributions
$\text{tr}(\rho_{\alpha}^{\mathrm{(in)}}\pi_{n})$ (blue curves). These are
generated for pure $|\alpha\rangle\langle\alpha|$ or mixed $\rho_{\alpha
}^{\mathrm{(in)}}$ and for $\pi_{n}$ reconstructed with the filter function or
without it. For all these options, the probability distributions (blue lines)
are so similar that they are indistinguishable on this scale.}%
\label{qfun_APD}%
\end{figure}
Care has to be taken that the optimisation problem is well conditioned in
order to find the true POVM of the detector. In finding their number basis
representation we are deconvolving a coherent state from our statistics which
is intrinsically badly conditioned due to the importance of the wings of the
Gaussian. Similar issues of conditioning have been discussed in the context of
state and process tomography, see, e.g., Refs.\ \cite{UnstableBoulant,
UnstableHradil}. Due to a large ratio between the largest and smallest
singular values of the matrices defining the quadratic problem, small
fluctuations in the probability distribution can result in large variations
for the reconstructed POVM. This can result in operators that approximate
really well the outcome statistics and yet do not exhibit a smooth
distribution in photon-number. We will discuss how to treat this problem in
the next section.

\begin{figure*}[ptbh]
\begin{center}
\includegraphics[width=13cm]{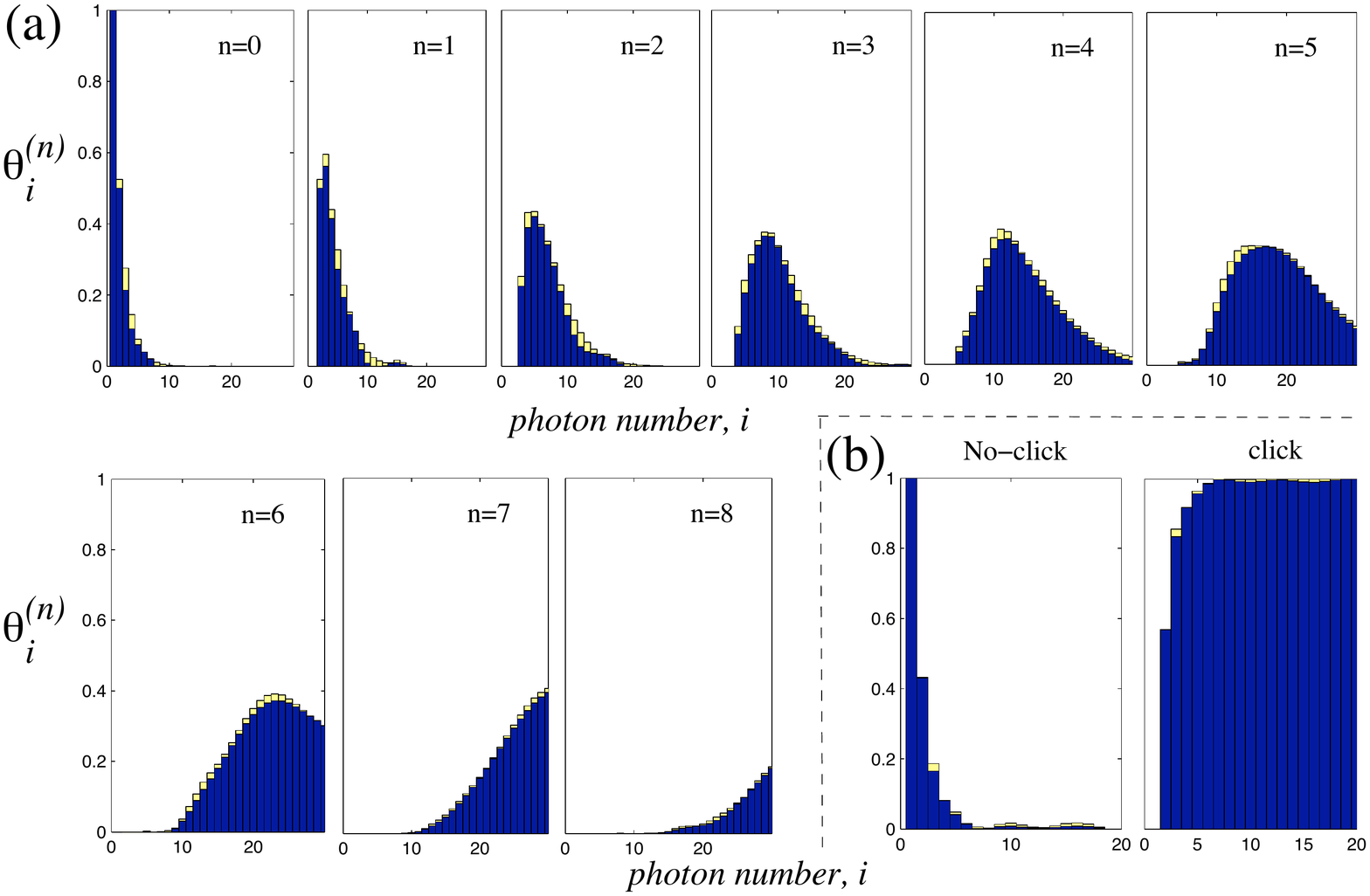}\\[0pt]
\end{center}
\caption{Reconstructed POVMs for (a) the photon-number resolving TMD and (b)
the APD ``yes/no'' detector. TMD POVMs were obtained up to element
$|60\rangle\langle60|$ (therefore $M=60$), but are shown up to $|30\rangle
\langle30|$ for display purposes. APD POVMs are shown in full. Stacked on top
of each $\theta_{i}^{(n)}$ where $n$ is the number of clicks we show
$|\theta_{i}^{n \mathrm{(\mathrm{rec})}} - \theta_{i}^{n \mathrm{(theo)}} | $
in yellow. ``rec'' stands for reconstructed and (theo) is the theoretical POVM
expected from {(a)} a TMD modelled with $3$ beam splitters of measured
reflectivities and 52.1\% overall loss {(b)} a theoretical APD with $43.2$\%
loss respectively. Note that this result was obtained with a regularized
optimisation as explained in next section . }%
\label{APDTMDPOVM}%
\end{figure*}
The measured probabilities for each outcome as a function of $|\alpha|^{2}$
are displayed in Fig.\ \ref{qfun_APD} and Fig.\ \ref{qfun_TMD}. The probability distributions (equivalent
modulo $1/\pi$ to the Q-function of the detector) show smooth profiles and
distinct photon number ranges of sensitivity for increasing number of
\textit{clicks} in the detector. Fig.\ \ref{APDTMDPOVM} shows the POVMs that
result from the optimisation in Eq.\ (\ref{optim}) which we will discuss
later. A first remarkable property is that $\pi_{n},$ being the POVM for $n$
clicks, shows zero amplitude for detecting less than $n$ photons. That is, the
detector shows essentially no dark counts. It should be noted that this was
not assumed at the outset and is purely the result of the optimization. This
sharp feature gives the detector its discriminatory power where $n$ clicks
means at least $n$ photons in the input pulse.

To assess the performance of our method we compare it to the model described
in the previous section. This time however, the BS used in the model are not
50/50 but its reflectivities ($R=[0.5018, 0.5060, 0.4192]$) were measured
experimentally. This was done measuring the reflected and transmitted beams of
a laser with a calibrated power meter. The yellow bars in
Fig.\ \ref{APDTMDPOVM}, show the absolute value of the difference between the
theoretical and the reconstructed POVM elements. The magnitude
\[
\Delta^{(n,i)}_{\theta} = |\theta_{i}^{(n, \mathrm{theo})} - \theta
_{i}^{(n,\mathrm{rec})}|
\]
is shown stacked on top of each coefficient of the POVM elements where
\textsl{\textrm{(theo)}} stands for theoretical and \textsl{\textrm{(rec)}}
for reconstructed. The small yellow bars reveal a good agreement with the
model. We also calculate a form of fidelity finding that
\[
F=\mathop{\rm Tr}{_{}}\left(  \left(  ({\pi_{n}^{\mathrm{\mathrm{(theo)}}}%
})^{\frac{1}{2}} \pi_{n}^{\mathrm{\mathrm{(rec)}}} ({\pi_{n}%
^{\mathrm{\mathrm{(theo)}}}})^{\frac{1}{2}}\right) ^{\frac{1}{2}}\right)
^{2}\geq98.7\%
\]
holds for all $n$.  Note that to calculate $F$ we normalized the POVM elements.
This overlap  indicates an excellent agreement between the two.

In addition, one can reconstruct a probability distribution: from the found
POVMs to fit the data. The reconstruction is plotted as dark blue bars in
Fig.\ \ref{qfun_TMD} and Fig.\ \ref{qfun_APD}. It is the equivalent of the Q-function had our probe states
$|\alpha\rangle\langle\alpha|$ with suitable complex $\alpha$ been strictly
pure. In fact, although formally distinct, the probability distribution
associated with the reconstructed POVM using mixed or pure states are
practically indistinguishable and are plotted together in Fig.\ \ref{qfun_TMD} for comparison.

\subsection{Detector Wigner functions}

An alternative representation of the detectors which can give us more insight
about their structure comes from the quasi-probability distributions such as
the Wigner Function \cite{UlfBook, Schleich-book}. Since the POVM elements
${\pi_{n}}$ are self adjoint positive-semi-definite operators, a Wigner
function $W_{n}$ can be calculated in the standard way from the POVM element
$\pi_{n}$:
\begin{equation}
W_{n}(x,p)=\frac{1}{\pi\hbar}\int_{-\infty}^{\infty}dy\,\langle x-y| {\pi}_{n
}|x+y\rangle e^{2ipy/\hbar},
\end{equation}
where we have, as usual, now identified $(x,p)\in\mathbbm{R}^{2}$ as phase
space coordinates of a single mode with $\alpha\in\mathbbm{C}$. However, since
the POVMs do not have unit trace, this detector Wigner function will not be
normalized,
\begin{equation}
\int_{-\infty}^{\infty}dx\,\int_{-\infty}^{\infty}dp\; W_{n }(x,p)<1.
\end{equation}
We should note that the marginals cannot be interpreted as probability
distributions but we can still use $W_{n}$ to calculate probabilities
according to
\begin{equation}
p_{\rho,n}=\mathop{\rm Tr}{_{}}\left( {\rho}{\pi}_{n }\right)  =\int_{-\infty
}^{\infty}dx\,\int_{-\infty}^{\infty}dp \; W_{\rho}(x,p)W_{n }(x,p).
\end{equation}

\begin{figure*}[ptb]
$%
\begin{array}
[c]{ccc}%
\includegraphics[width=8cm]{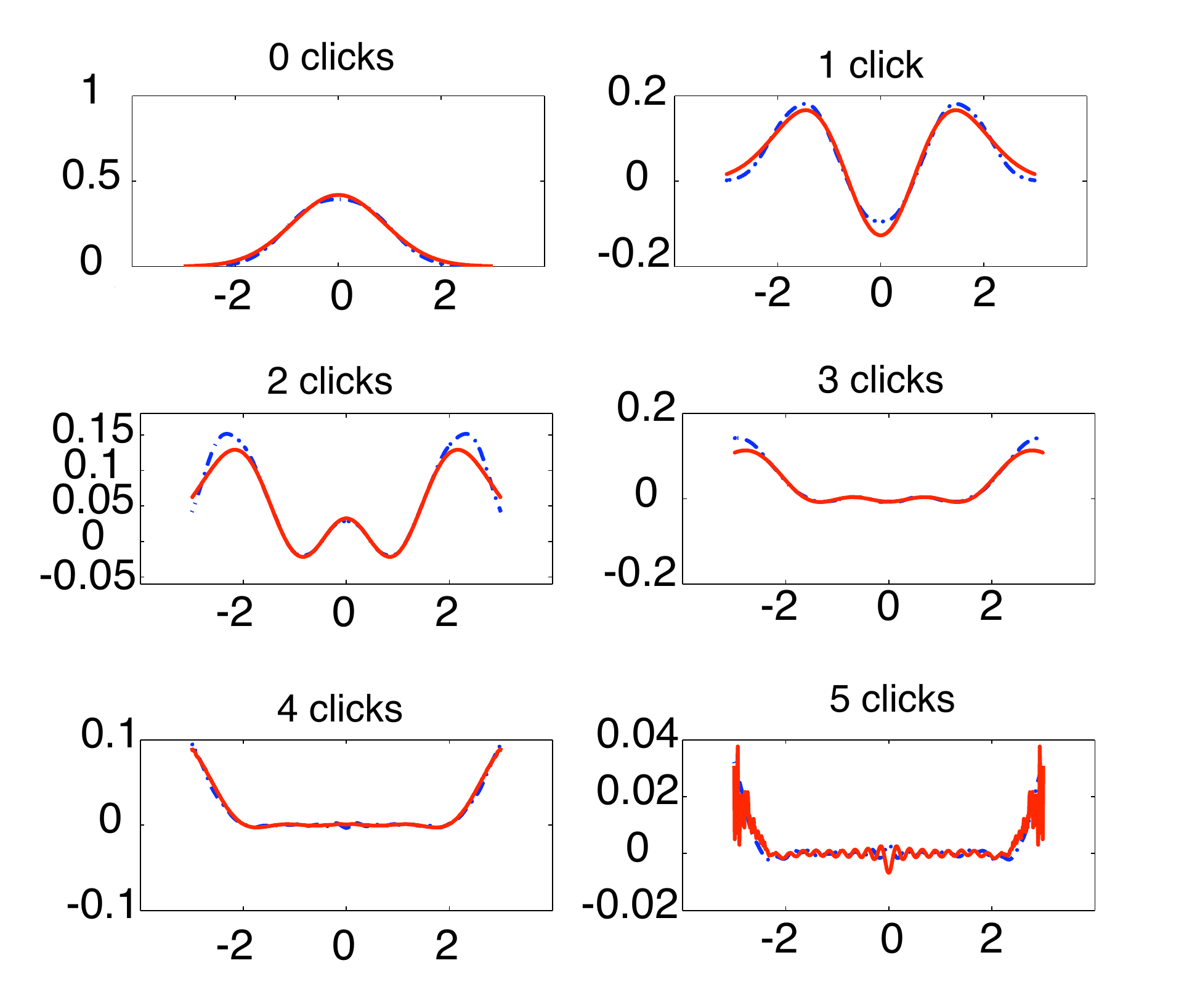} &
\vline &
\includegraphics[width=8cm]{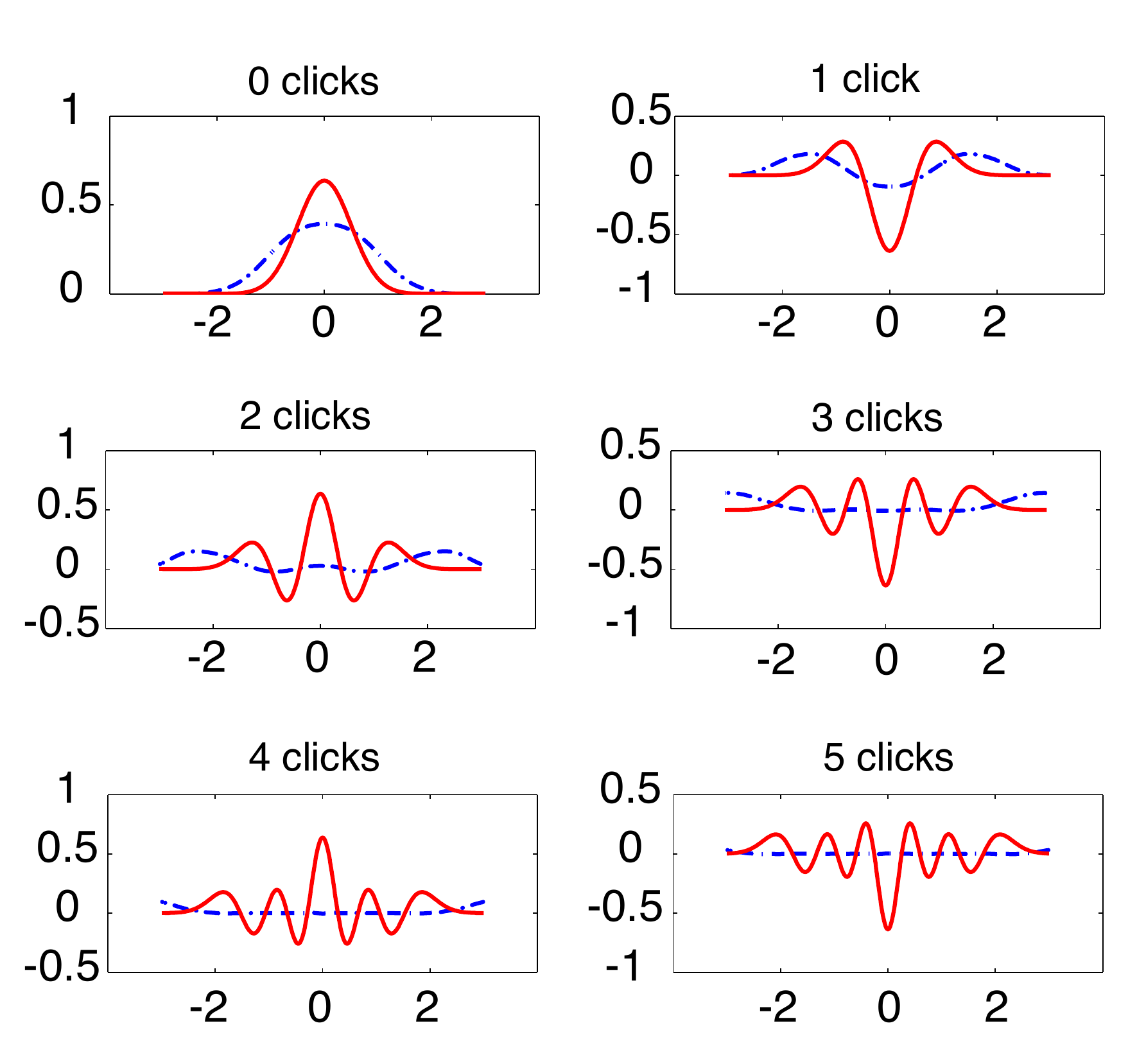}\\
(a) &  & (b)\\
&  &
\end{array}
$\caption{Wigner function of the first POVM elements of the TMD. Since the
detectors have no phase reference, their Wigner functions are rotationally
symmetric with respect to their center and a cut contains all the information.
The dotted blue curve represents the Wigner function of the reconstructed
POVMs from $0$ to $5$ clicks. In red we can see the theoretical Wigner
function for: (a) a theoretical TMD with 52\% loss. (b) a theoretical TMD
without loss. Paying attention to the scale we observe how dramatic the effect
of loss is at damping the ripples in the Wigner function. It is also worth
noting that the end ripples in (a) for the ``$5$ click'' are just an edge
effect due to number state cutoff.}%
\label{wigner2}%
\end{figure*}

Since none of the detectors have phase sensitivity their Wigner functions are
rotationally symmetric around the origin. In Fig.\ \ref{wigner2} we display a
cut of the TMD Wigner function for the following POVM elements:
\[
\{ \pi_{0}, \pi_{1}, \pi_{2}, \pi_{3}, \pi_{4}, \pi_{5}\}.
\]
Higher clicks are not displayed because their amplitude is too small to be
compared with the rest. The interesting feature about the plot is the
comparison with the theoretical TMD Wigner functions one can generate with the
model. Indeed, comparing a theoretical loss-less TMD with the measured one we
see how the amplitude of the Wigner function decreases rapidly for higher
photon numbers. On the other hand, comparison with the lossy theoretical model
reveals a good agreement.

\section{Ill Conditioning and regularisation}


One of the main problems encountered in the tomographic characterisation of
the detectors has to do with the numerical stability of the reconstruction.
Such problems are common in tomography \cite{UnstableBoulant, UnstableHradil}.
Consider for example the transformations involved in the inverse Radon
transform and their inherent instabilities. Note also how going from the
Q-function to the P-function is not always well defined \cite{Schleich-book}.
Multiple tools exist to bridge the link between homodyne tomography and the
density matrix description \cite{lvovsky-pattern}. One of them involves the
use of pattern functions \cite{dariano-pattern, leonardt-pattern,
wunsche-1997}. That is, finding some functions $G_{k}(\alpha)$ such that
\[
\int{\ Q^{(n)}(\alpha) G_{k}(\alpha) d^2\alpha} = \theta_{k}^{(n)}.
\]
However, finding the appropriate $G_{k}$ involves the irregular wave functions
(particular unbounded solutions of the Schr\"odinger equation) and proving
them to be appropriate is typically as hard as estimating the error
\cite{pattern-fun}. The use of maximum likelihood has also been explored and
particularly for detector tomography \cite{dariano-2004-93,
PhysRevA.64.024102}. However, the speed of the convergence is not generally
guaranteed to be high, becoming exponential for certain problems. Our
approach, following the spirit of maximum-likelihood, translates the problem
into a quadratic optimisation one allowing for efficient semi-definite
programming (SDP) (cf.\ Eq.\ (\ref{optim})). We discuss here the details,
approximations and filters that lead to our solution of the problem.

\subsection{Truncating the Hilbert space}

The data was measured up to $|\alpha|^{2} = 150$ but was truncated at lower
values in phase space.  This was done to avoid noisy behavior and the emergence of new regimes in the behavior of the detector. Memory effects requiring a larger POVM space were thus avoided as discussed in the experimental section.  Notably, the effects related to the detector's dead time, after-pulsing or the dark counts from possible over heating were avoided staying in a low ($|\alpha|^{2} \simeq 70$) photon number regime.

\subsection{Pure vs.\ mixed}

The Q-function of our detector (directly measured) is proportional to
\begin{equation}
\label{probability}p_{\alpha,n}=\mathop{\rm Tr}{_{}}\left( |\alpha
\rangle\langle\alpha| \pi_{n}\right)  .
\end{equation}

From Eq. (\ref{matrix_equation}) and Eq. (\ref{simple_optim}), for a diagonal
POVM, we can write the problem as

\begin{figure}[ptb]
\includegraphics[width=8cm]{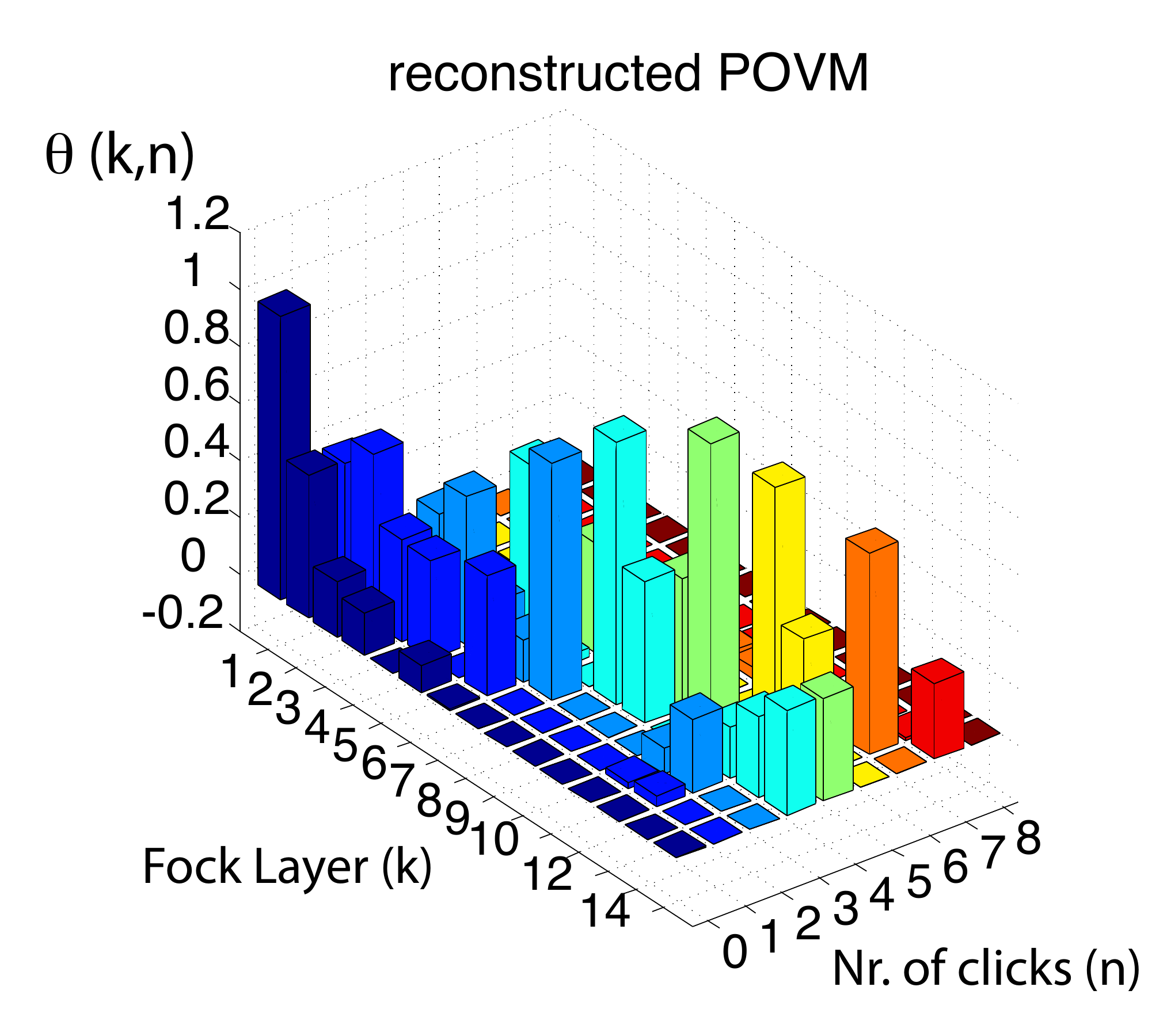}\caption{POVM
reconstruction, using only minimisation from Eq.\ (\protect\ref{directmin}). Dark
blue: $\pi_{0}=\sum_{i=0}^{15}\theta_{i}^{(0)}|i\rangle\langle i|$, lighter
blue, $\pi_{1}=\sum_{i=0}^{15}\theta_{i}^{(1)}|i\rangle\langle i|$, etc.}%
\label{badpovm}%
\end{figure}
\begin{equation}
\min{\Vert{P}-F\Pi\Vert_{2}}.\label{directmin}%
\end{equation}
with the usual constraints $\pi_{n}\geq0$ and $\sum_{n}\pi_{n}=\mathbbm{1}$.
Using a semi-definite solver such as Yalmip, the obtained POVMs $\{\pi_{n}\}$
shows irregular dips and a structure quite dissimilar from what a TMD is
expected to do.
\begin{figure}[ptb]
\includegraphics[width=8cm]{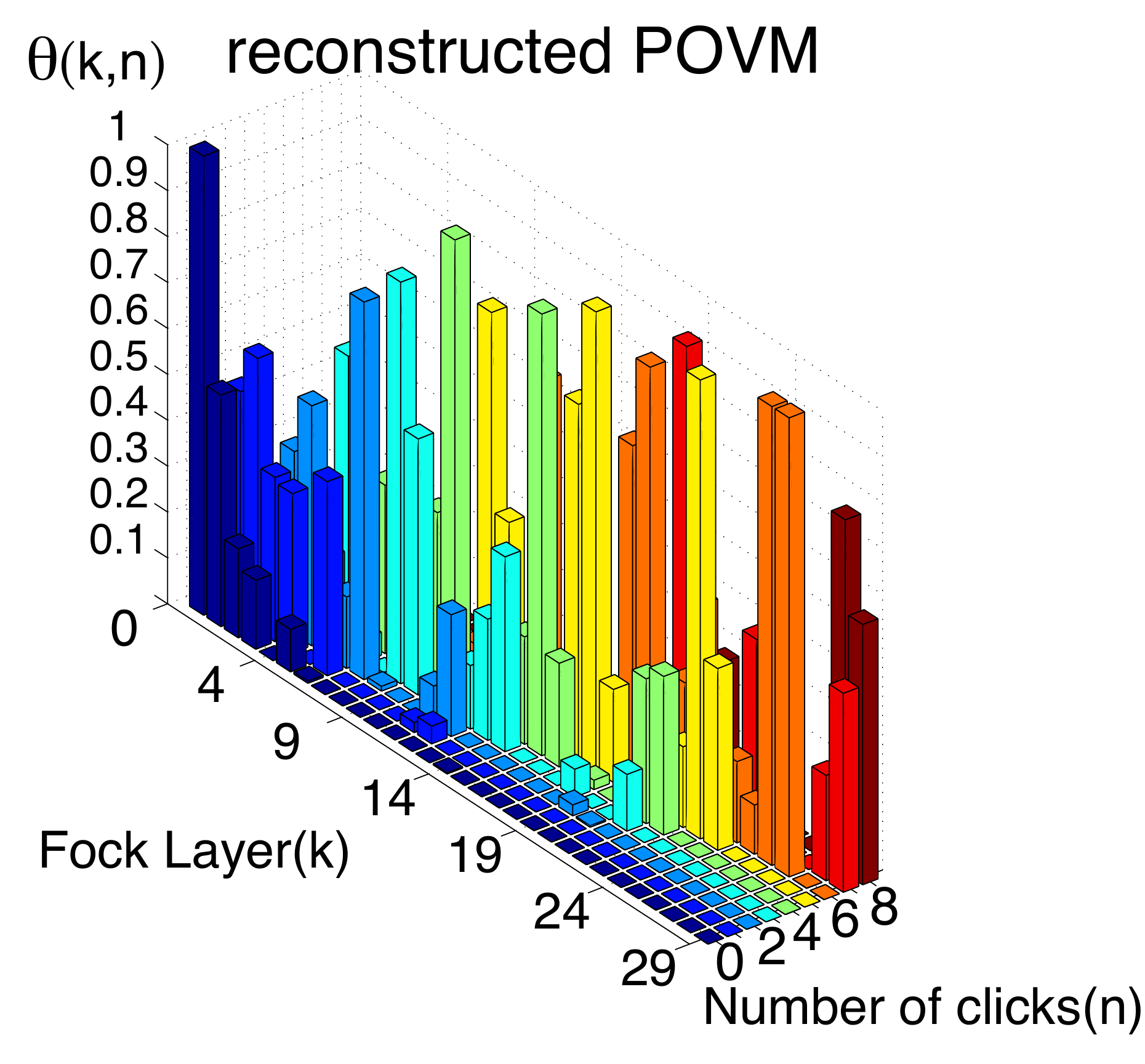}\caption{POVM
reconstruction, using only minimisation from Eq.\ (\protect\ref{directmin}). Dark
blue: $\pi_{0}=\sum_{i=0}^{15}\theta_{i}^{(0)}|i\rangle\langle i|$, lighter
blue, $\pi_{1}=\sum_{i=0}^{15}\theta_{i}^{(1)}|i\rangle\langle i|$, etc,
displayed up to number layer $30$.}%
\label{longbadpovm}%
\end{figure}
The Fig.\ \ref{badpovm} shows a typical result, and
Fig.\ \ref{longbadpovm} shows it for higher photon numbers revealing an even
more irregular structure. 

\subsubsection{Describing the laser's amplitude uncertainty}

\begin{figure}[ptb]
\includegraphics[width=8cm]{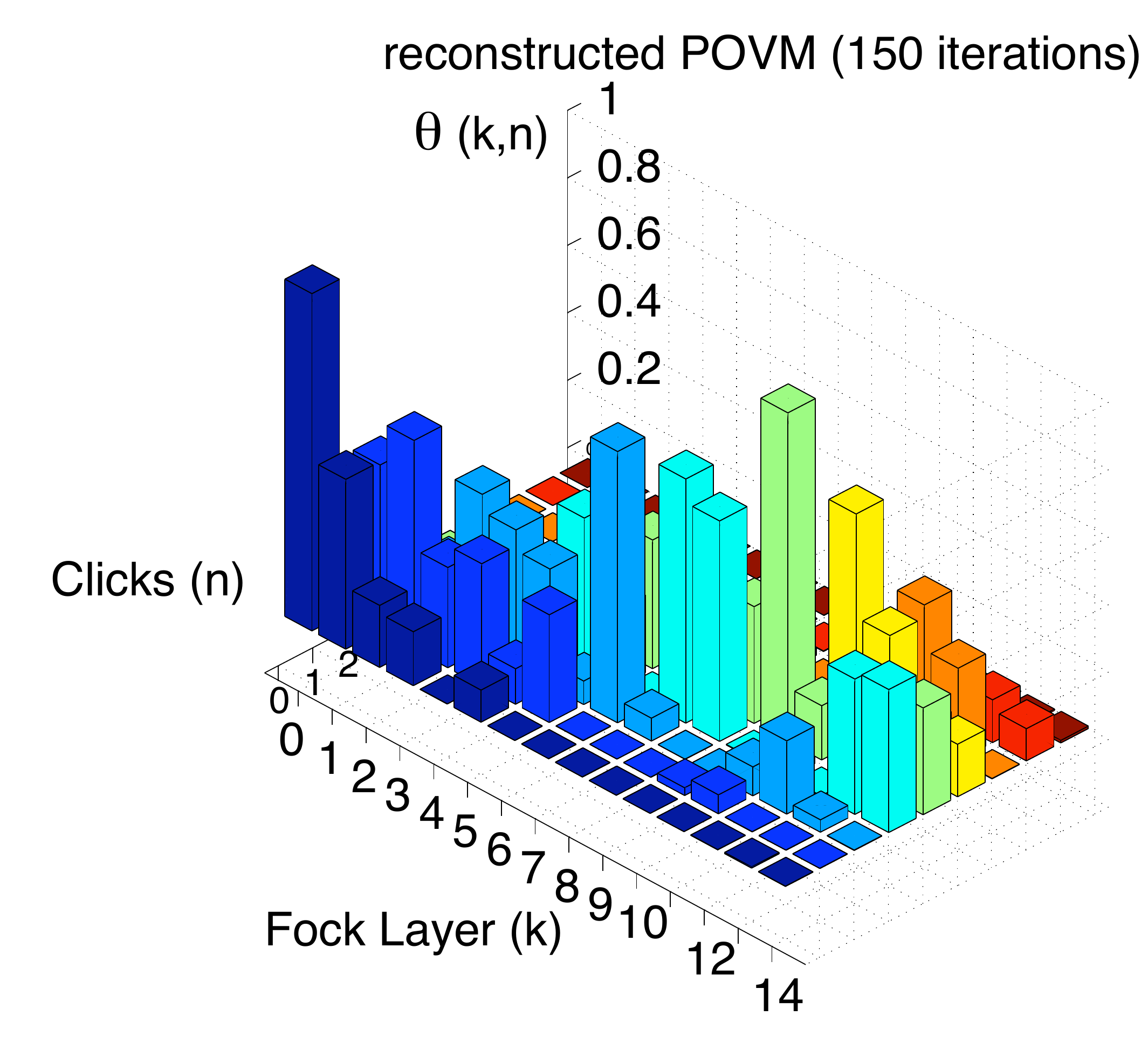}\caption{POVM
reconstruction, using direct averaging (150 runs with 1\% Gaussian noise).}%
\label{KennyAve}%
\end{figure}

\begin{figure}[ptb]
\includegraphics[width=8cm,angle=0]{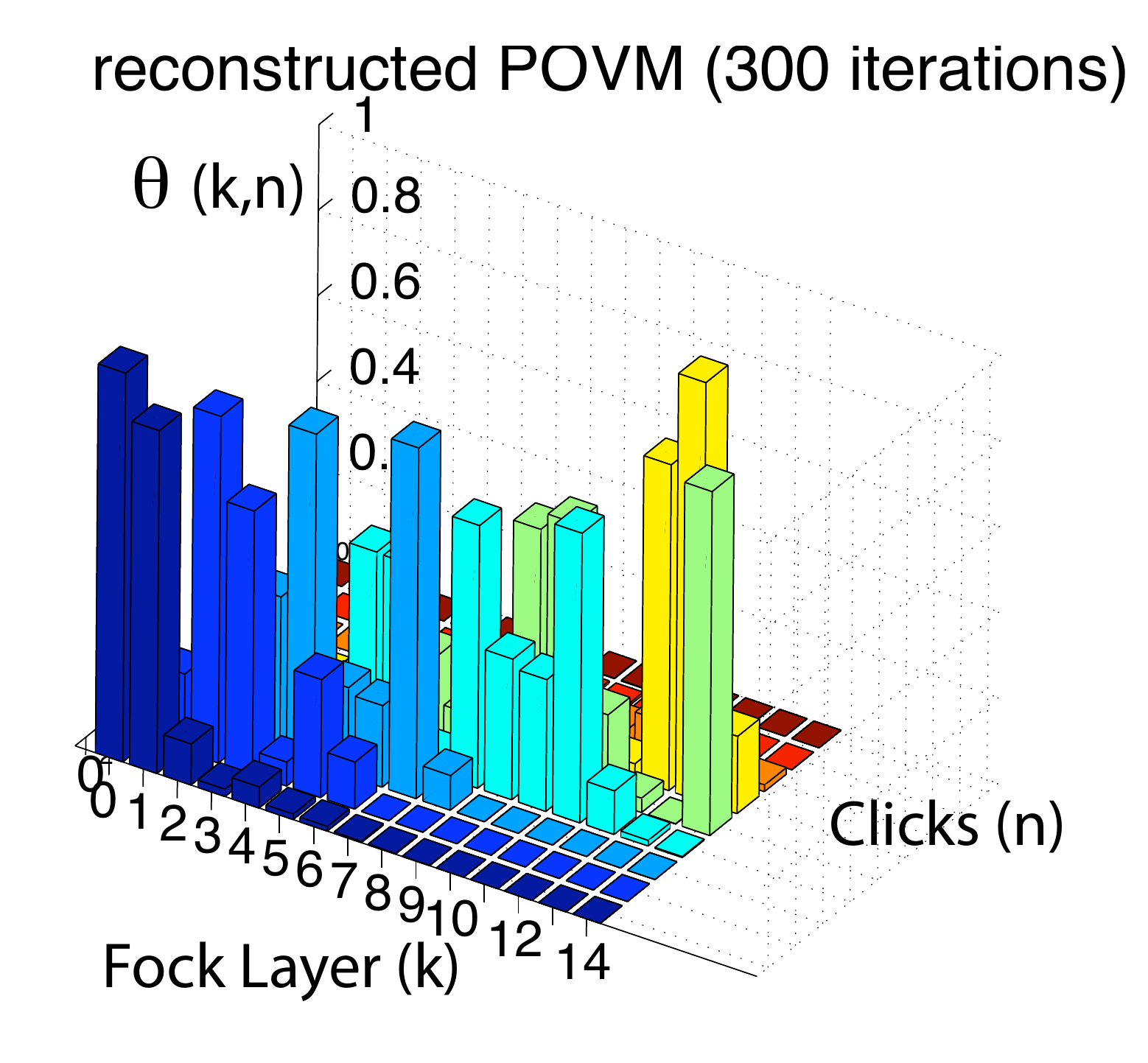}\caption{POVM
reconstruction, using direct averaging (300 runs with $2$\% Gaussian noise).}%
\label{KennyAve300}%
\end{figure}

A first meaningful observation is that some level of uncertainty existed in the
 amplitude $x=|\alpha|^{2}$ of the coherent states. If $D$ values of $x$ were measured then the real
\[
\bar{x}=(x_{1},x_{2},,\dots,.,x_{D})
\]
might actually have been
\[
\bar{x}_{{\delta}}=\left(  x_{1}(1+\delta_{1}),x_{2}(1+\delta_{2}),\dots
,x_{D}(1+\delta_{D})\right)  ,
\]
with some vector of errors $(\delta_{1},\dots,\delta_{D})$. To address
the effect of this uncertainty on our minimisation we can artificially
introduce noise and then average over many runs of the optimisation.
In other words, since $F$ in Eq.\ (\ref{directmin}) depends on the measured
values of $|\alpha|^2$, we
can substitute $\bar{x}$ with $\bar{x}_{{\delta}}$ where
$\delta=(\delta_{1},\dots,\delta_{D})$ are independent and identically
distributed random variables.  Using $\bar{x}_{{\delta}}$ we run the
optimisation and obtain a family of estimated POVM elements (each element
of the family corresponds to a run of the optimisation with a different $\delta$)
As a first approximation we may use a Gaussian probability distribution with zero mean and
$\sigma=2\%|\alpha|^{2}$.  Note that $2$\% was the measured variance of the laser amplitude
from pulse to pulse as shown in Fig.\  \ref{Pulse_Jitter}.
Subsequently we average
over the POVMs obtained with different \textquotedblleft
jitters\textquotedblright\ ${\delta}$ in $N$ realizations, obtaining
\[
\pi_{n}^{\mathrm{(average)}}=\sum_{j}{\pi_{n}^{({\delta}_{j})}}/N.
\]
$200$ iterations of the optimisation with subsequent averaging
improves the appearance of the POVMs but barely solves the
\textquotedblleft dips\textquotedblright\ observed. Fig.\
\ref{KennyAve} and Fig.\ \ref{KennyAve300} are an example of this
approach, showing that this kind of averaging does 
not properly
counteract the fluctuations in the reconstructed POVM.

\subsubsection{Using mixed input states}

\begin{figure}[ptb]
\includegraphics[width=9cm,angle=0]{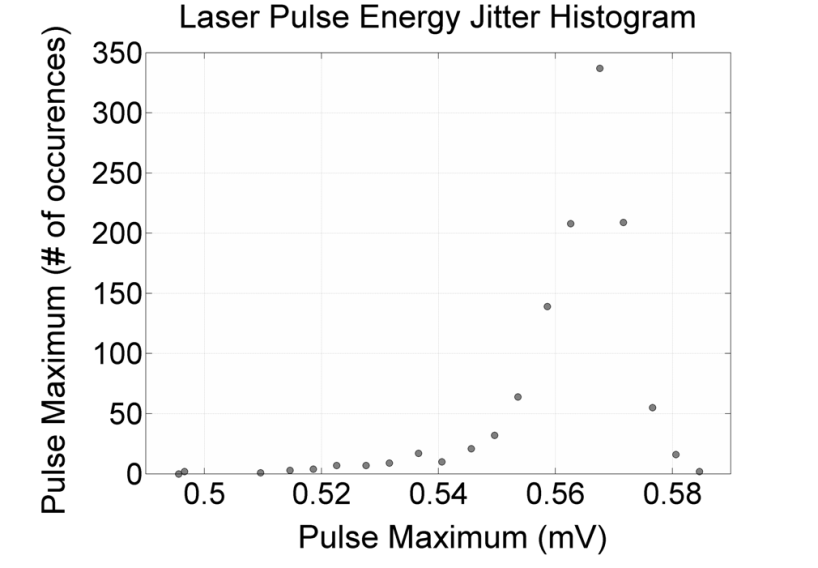}
\caption{
Measurement of the laser's amplitude variations from pulse to pulse.
}%
\label{Pulse_Jitter}%
\end{figure}

A key obsevation showing that the previous approach is not the appropriate
treatment of uncertainty in $x$ is that each probe state would be best
described by a mixture of coherent states,
\begin{align}
\rho_{x}  & =\int d^{2}\beta|\beta\rangle\langle\beta|f_{x}(\beta)\\
& =\sum_{n,m=0}^{\infty}R_{n,m,x}|n\rangle\langle m|.
\end{align}
Here, $f_{x}$ would be some distribution centered around $x$ in phase space, leading
to a mixed Gaussian state in case of a Gaussian classical probability
distribution. We can integrate this state $\rho_{x}$ over the complex phase
since we have no phase reference available and focus solely on the amplitude
of the coherent states or mixtures thereof. Measurements reveal that the
intensity of the laser varies from pulse to pulse following a distribution
that looks like a Lorentzian with a tail (see Fig.\  \ref{Pulse_Jitter}).
A good approximation can however be made using a Gaussian distribution,
leading to a Gaussian state, with standard deviation $\sigma=0.02|\alpha|^{2}%
$, implying
\[
R_{n,m,\alpha}=\frac{1}{\sigma\sqrt{2\pi}\sqrt{n!m!}}\int\beta^{n+m}%
e^{-\beta^{2}}\;f_{x}(\beta)\;d\beta.
\]
with $f_{x}(\beta)=e^{-(\beta^{2}-x)^{2}/(2\sigma^{2})}$. The detection
probability for outcome $n$ is then
\begin{equation}
p_{n}(\alpha)=\sum_{k=0}^{\infty}R_{k,k,\alpha}\theta_{k}^{(n)}%
.\label{probmix}%
\end{equation}
To simplify these calculations we can write a distribution in $\sqrt
{x}=|\alpha|$,
\begin{equation}
\rho_{|\alpha|}=\int d^{2}\beta|\beta\rangle\langle\beta|\;g_{|\alpha|}%
(\beta)\label{mixedstate}%
\end{equation}
with
\[
g_{\alpha}(\beta)=e^{-(\beta-\alpha)^{2}/(2\Gamma^{2})}.
\]
In this case $\Gamma$ has been chosen such that the approximation $f_{x}%
(\beta)\simeq g_{\alpha}(\beta)$ holds. These subtleties however do barely
alter our results and POVMs are as irregular as previously.

To evaluate the difference introduced by the pure ($|\alpha\rangle
\langle\alpha|$) or mixed state ($\rho_{| \alpha|}$) approach we have studied
their influence on the reconstructed POVMs. In the regularised optimisation
(i.e., for our final results), we have compared the POVMs obtained with each
description finding that
\begin{equation}
\frac{\|\Pi_{\mathrm{pure}}-\Pi_{\mathrm{mixed}}\|_{2}}{\|\Pi_{\mathrm{mixed}%
}\|_{2}}\leq0.7\%
\end{equation}
and the largest relative difference between any two $\theta_{k}^{(n)}$ coming
from a mixed state or a pure state derivation was $1.3\%$. Furthermore, the
reconstructed probability distributions are so close that they are
indistinguishable on the scale of Fig.\ \ref{qfun_TMD}. This reinforces our
earlier expectation that technical noise in the laser will be negligible when
using single-photon-level coherent states. This differs from homodyne
tomography where technical noise can shift a strong local oscillator to a
nearly orthogonal state.

However, since the problem of the irregular POVMs is not solved by the mixed
state description we need to look further into the origin of these
irregularities. One first remarkable (but expected) property is that large
variations in the photon number degree of freedom of the POVMs result in
minuscule differences in the probability distributions (see
Fig.\ \ref{APDTMDPOVM}). Since one convolutes the photon number distribution
with a Gaussian in $\alpha$ to obtain the Q-function this behavior has been
expected. Conversely this means that small errors or statistical fluctuations
in the Q-function can result in large errors in the POVM elements. Consider
for example that if instead of
\[
\min{\|{P} - F \Pi\|_{2} }%
\]
we try to minimise
\[
\min{\| F^{-1} {P} - \Pi\|_{2}}%
\]
the SDP solver finds no sensible solution. This is because using the
Moore-Penrose pseudo-inverse we find $F^{-1} F \neq\mathbbm{1}$ due to its
inherent ill conditioning, meaning that the ratio of largest and smallest
singular values in $F$ is large.

Various methods exist to try and regularise these problems. Whatever the
chosen method it should assume as little knowledge as possible about the
specific form of the sought POVM. Since $F$ has very small values
for high photon numbers one could enhance those values while preserving the
minimisation target. For example we could run the optimisation
\begin{align*}
\min{\|{P} D - F \Pi D\|_{2} }%
\end{align*}
where $D$ is a diagonal matrix aimed at regularising the problem. This can be
shown to introduce some improvement but does not solve the ``dip problem''
completely. It is also hard to find the exact form of $D$ that yields ``good''
results without any prior knowledge about the expected POVMs. In addition
(roughly speaking) it is hard to find a balance between having good results
for low photon numbers and for high photon numbers.

\begin{figure}[ptbh]
\begin{center}
\includegraphics[width=8cm,angle=0]{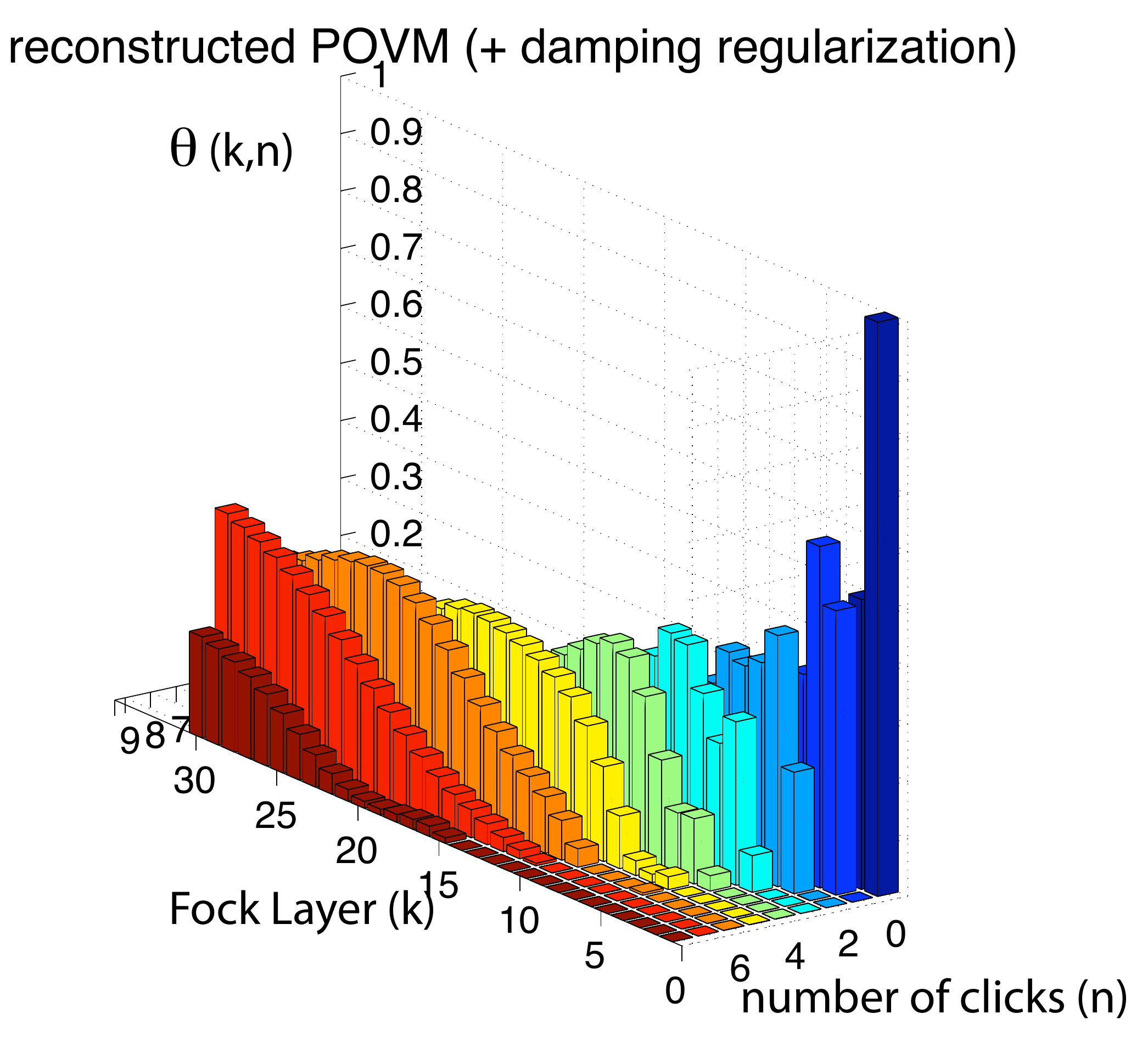}
\end{center}
\caption{Minimisation using damping method on Eq.\ (\protect\ref{damping}). Note that
the point of view is opposite that of the previous plots. We see some dips
around the $5$-th and $7$-th number layer.}%
\label{dampingfigure}%
\end{figure}Another approach is to introduce a sort of damping or specific
penalisation. For example one could define a diagonal matrix $M$ such that
\[
M_{i,j}=\delta_{i,j} /j,
\]
and use it to redistribute the weight of each POVM element, avoiding
unreasonably large POVM element amplitudes (that compensate for low values in
$F$). The optimisation could be recast as,
\begin{equation}
\label{damping}\min{\{\|{P} - F \Pi\|_{2} +0.03 \| M \Pi\|_{2} \}}.%
\end{equation}
A result of this can be seen on Fig.\ \ref{dampingfigure}. This method has the
same shortcomings as the previous one: it is sensitive to the choice of
parameters and the exact form of $M$ is hard to determine without detailed
prior assumptions.

A more reasonable method is to capture the relative smoothness of the POVM
from a lossy detector. This method is also called smoothing regularisation
\cite{Convex}. In this case one single assumption needs to be made. The POVMs
should exhibit a certain degree of ``smoothness''.

\subsection{Smooth or not?}

Let us first define what we mean by smooth. Smooth will mean in this context
that the difference $|\theta_{k}^{(n)} - \theta_{k+1}^{(n)}|$ is small for all
$k$ and $n$. In the optimisation context we will mean that our minimisation is
defined as follows:
\begin{equation}
\label{SDP-optim}\min{\{\|{P} - F \Pi\|_{2} + y {S}\}}%
\end{equation}
with
\[
{S}=\sum_{k,n}( \theta_{k}^{(n)}-\theta_{k+1}^{(n)})^{2}
\]
for some fixed value of y.  The smoothing function $S$ will be independent of the detector, and will mildly penalize non-smooth POVM elements. This approach is further
substantiated by the observation that the resulting POVMs are largely
independent of the weight $y$ that is given to the smoothness penalty.

As most quantum detectors, especially those disucussed here are lossy, this is a particularly plausible
feature. Indeed, if an optical detector has a POVM element with non-zero
amplitude in $|n\rangle\langle n|$, then if it is lossy, it will have a
positive amplitude in $|n+1\rangle\langle n+1|$, 
$|n+2\rangle\langle n+2|,
\dots, |n+K \rangle\langle n+K| $,  
decreasing with $K$ but different from zero. In fact, in general, if the
detector has a finite efficiency $\eta$ which can be modelled with a BS, it
will impose some smoothness on the distribution $\theta_{k}^{(n)}$. That is
because if $G(k)$ is the probability of registering $k$ photons and
$H(k^{\prime})$ is the probability that $k^{\prime}$ were present, then the
loss process will impose \cite{kenny-thesis}:
\[
G(k)= {\displaystyle\sum\limits_{k^{\prime}}} \binom{k^{\prime}}{k}\eta
^{k}(1-\eta)^{k^{\prime}-k}H(k).
\]
Consequently, if $\theta_{k}\neq0$, then $\theta_{k+1}$, $\theta_{k+2}$ etc.
cannot be zero, but will have some relatively smooth distribution. This simple
physical argument makes a certain smoothness plausible (but still should allow
sharp transitions for $m < n$).

For this detector (and for any photodiode based detector) assuming loss is
reasonable and can make the \textquotedblleft smoothness\textquotedblright%
\ requirement appropriate. Let us however see if, without looking at
the specific shape of our POVM, we can find an optimal
smoothing coefficient $y$ and justify further the use of the
smoothing regularisation.

\begin{figure}[ptbh]
\begin{center}
\includegraphics[width=8cm]{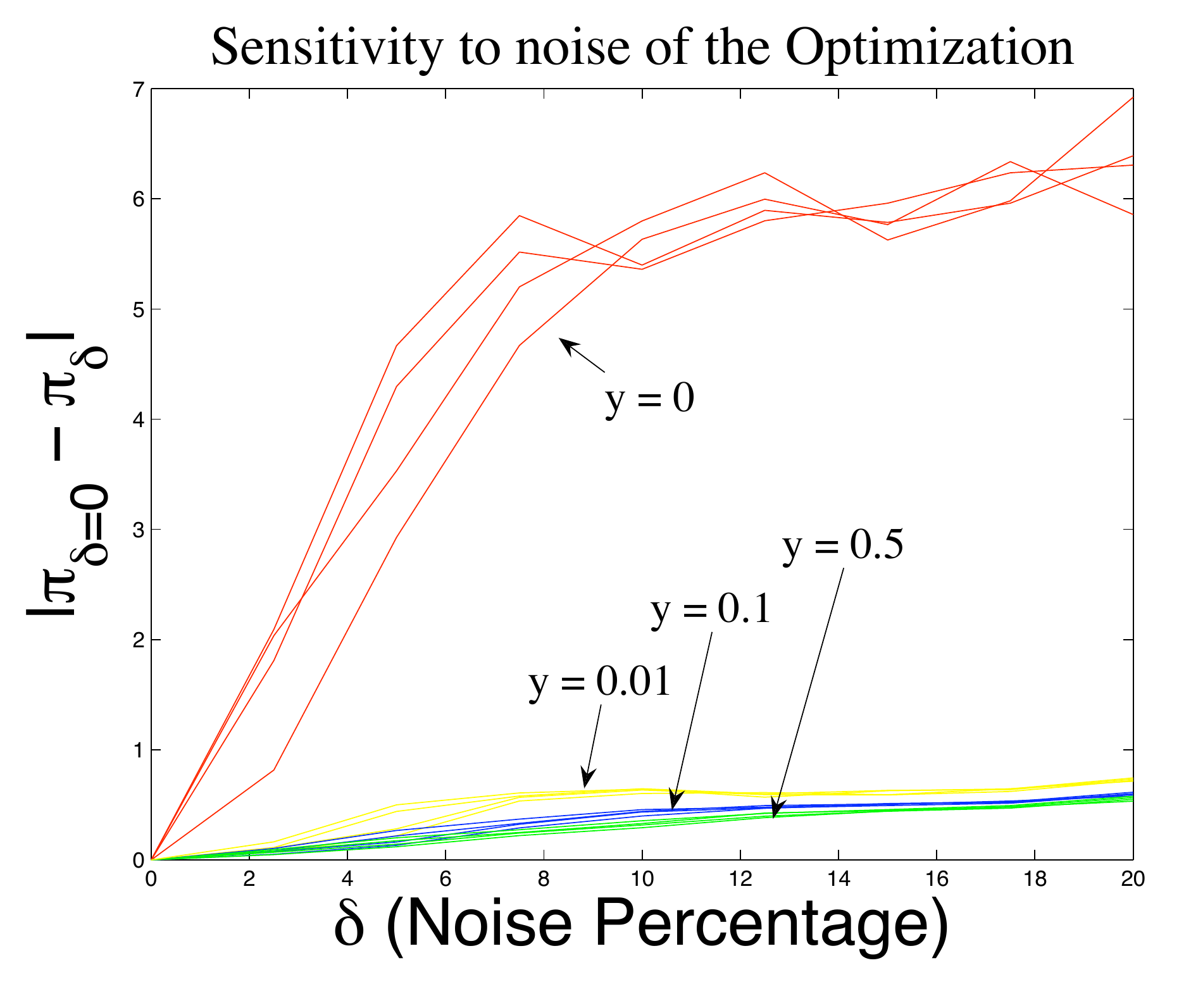}
\end{center}
\caption{Illustration of the sensitivity to noise for two different
minimisation methods ($y=0$ corresponds to no regularisation, and $y \neq0$ to
an approach using a smoothing regularisation). For each value of $y$ and
$\delta$ we have run the optimisation $4$ times and displayed the results here
to illustrate this variation.}%
\label{deltanoise}%
\end{figure}

\begin{figure}[ptbh]
\begin{center}
\includegraphics[width=9cm]{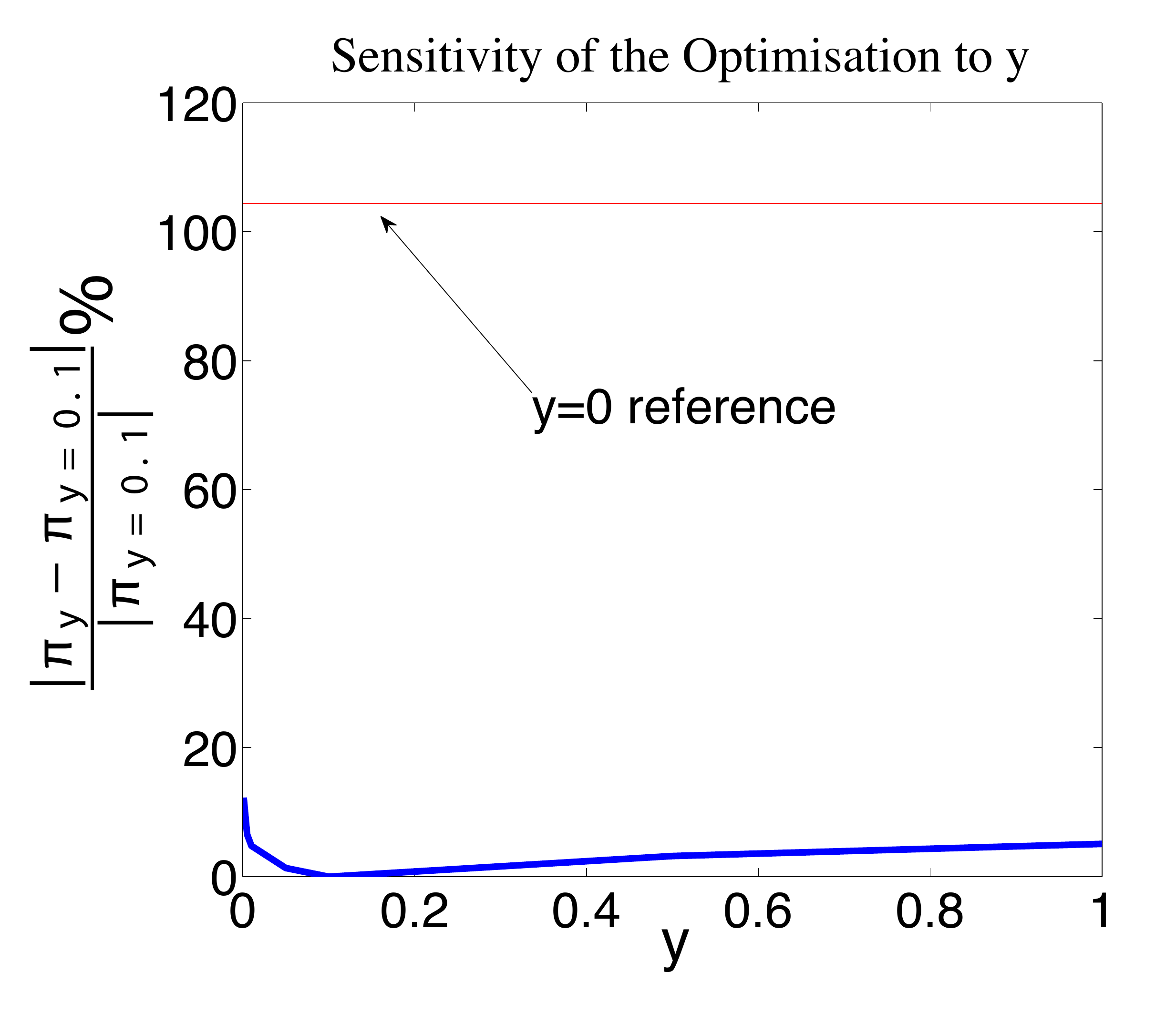}
\end{center}
\caption{Illustration of how sensitive the optimisation is to the specific
choice of $y$. This plot shows the relative error with respect to the POVM
elements obtained using $y=0.1$, as a function of $y$. In red, and only for
reference (since it does not change with $y$), the value of the relative error
for $y=0$ (no smoothing) is shown. We vary $y$ in the range $[0.001,1]$. It is
remarkable that a $10000$\% variation in $y$ results in only a $12$\%
variation. For $y \in[0.05,0.2]$ the relative error is less than $2$\% in
$\Pi$. }%
\label{ysensitive}%
\end{figure}
One way to test this method is to quantify how resilient it is to noise in the
data. To do so we introduce additional noise in $x = |\alpha|^{2}$ to the
measured data. For example, we can alter $x$ in ${P}_{i,n} = P(x_{i} (1 +
\delta_{i}),n)$ where $\delta=(\delta_{1},\dots, \delta_{D})$ is again a
vector of random variables distributed with a Gaussian distribution with zero
mean. This simulates a statistical uncertainty in the measurement of the
coherent state. To see its effect on the reconstruction we use the figure of
merit $\| \Pi_{\delta}- \Pi_{\delta=0} \|_{2}$. This quantity should evaluate
how POVMs differ from the one without noise. It is seen that the additional
smoothing penalty makes the optimisation more robust, largely independent of
the value of $y$ (we can multiply $y$ by a $100$ and stay in the same regime).
Using this smoothing regularisation with noisy data seems therefore a good
choice.

We have seen how smoothing makes the optimization more robust against
noise but we should also ask how sensitive this optimisation is to the
exact choice of $y$. To do so we may use the following procedure:
compare the POVM obtained using $y=0.1$ with that obtained varying $y$
over $4$ orders of magnitude. In Fig.\ \ref{ysensitive} we plot the
relative error $\displaystyle{\
100\ast{|\Pi_{y}-\Pi_{y=0.1}|}/{|\Pi_{y=0.1}|}}$. Remarkably doubling
the value of $y$ results in an overall relative error in the POVM of
less than $1$\%. Multiplying (or dividing) $y$ by 10 gives a variation
below $5$\% and $100$-fold variation results in a $12$\% variation. If
we compare how this differs from the $y=0$ case which is $110$\%
different then we can conclude that the optimisation is quite
insensitive to the exact choice of the \textit{smoothing parameter}
$y$. The following table provides some values for reference.

\begin{center}%
\begin{table}
\begin{tabular}
[c]{c|c|c}%
$y$ & $y$ variation & $\Pi$ relative error\\\hline
0.0001 & x/1000 & 27.3\%\\
0.001 & x/100 & 12.2\%\\
0.01 & x/10 & 4\%\\
0.05 & x/2 & 1\%\\
0.5 & x 5 & 3\%\\
1 & x 10 & 5\%\\
&  &
\end{tabular}
\caption{This table illustrates how sensitive the optimisation
is to the particular choice of $y$ (the reference smoothing strenght is $y=0.1$)}
\end{table}
\end{center}

\begin{figure*}[ptb]
\includegraphics[width=15cm]{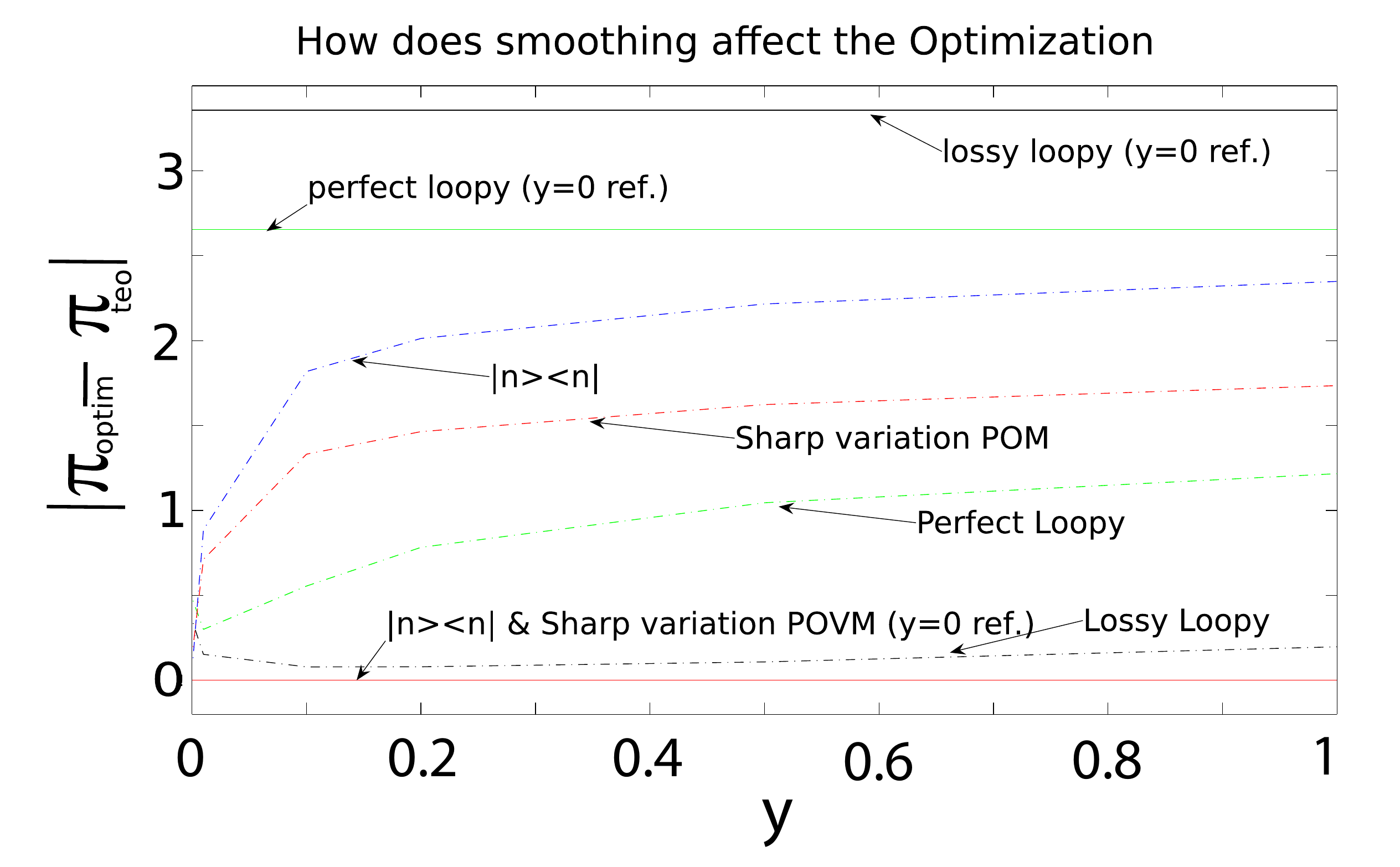}
\caption{Illustration of how too much smoothing can fail to capture the sharp
variations of a POVM. We define $\Pi_{\mathrm{theo}}$ as the matrix containing
the POVM elements of the theoretical POVMs. From them we generate a
probability distribution and reconstruct the POVMs $\Pi_{\mathrm{optim}}$ with
the smoothing regularised optimisation. The dotted lines represent
$\|\Pi_{\mathrm{optim}} - \Pi_{\mathrm{theo}}\|_{2}$ for different values of
$y$ and for a variety of POVMs (see following plots). The horizontal lines
represent that same difference for $y=0$ and are plotted for reference.}%
\label{ynoise}%
\end{figure*}

\subsection{Sharp and smooth}

There is of course a limit to how much we can penalize non-smooth POVMs. Is it
possible for the smoothing regularisation to wash out all the sharp features
of the POVM, thus smoothing in excess? This of course is a legitimate question
that further restricts the reasonable range for $y$. To study that effect we
analyse four cases:

\begin{enumerate}
\item A theoretical loss-less TMD, based on the model described in
Eq.\ (\ref{recursive-loopy}).

\item A lossy TMD, based on the above with added loss from an $R=52\%$ BS.

\item A perfect photon number detector, that is with $\pi_{n} =
|n\rangle\langle n|$.

\item An artificial POVM with sharp variations, containing the POVM
elements:  
\begin{eqnarray*}
\pi_{0}&=& |0\rangle\langle0| + |2\rangle\langle2| \\
\pi_{1} &=& |1\rangle\langle1| +1/2 |3\rangle\langle3| \\
\pi_{2} &=& 1/2 |3\rangle\langle3| + |4\rangle\langle4| + |5\rangle\langle5| \\ 
\pi_{3}&=& |7\rangle\langle7|\\
\pi_{4}&=& 1/4 |6\rangle\langle6| + 1/4 |8\rangle\langle8|\\ 
\pi_{5}&=& 1/4 |6\rangle\langle6| + 1/4 |8\rangle\langle8|\\ 
\pi_{6}&=& 1/2 |6\rangle\langle6|\\
\pi_{7}&=& 1/2 |7\rangle\langle7|\\ 
\pi_{8} &=& 1/2 |8\rangle\langle8|+  \sum_{k=9}^{60} | k \rangle\langle k |
\end{eqnarray*}
and observing
\[
\sum_{i} \pi_{i} = \mathbbm{1} .
\]

\end{enumerate}

To study the smoothing we generate the POVM elements $\{\pi_{n}\}$
numerically, build a probability distribution $\mathop{\rm Tr}{_{}}\left(
\rho_{\alpha} \pi_{n}\right)  $ and retrieve the $\pi_{n}$ using the
optimisation from Eq.\ (\ref{SDP-optim}) for an increasing range of $y$. Then
we compare these results with the theoretical POVMs we defined in order to
generate the PD. All optimisations are done using the mixed-state approach
from Eq.\ (\ref{mixedstate}). Broadly speaking, we find two distinct
behaviors: POVMs with terms that decay slowly in photon number need
regularisation and are quite insensitive to the precise $y$. For sharp POVMs
(without loss) the range $0<y<0.01$ preserves their shape quite well, but
further smoothing hides their true shape. These properties are further
illustrated in the figures that follow.


\subsubsection{Lossy TMD}
\begin{center} 
\begin{figure*}
\includegraphics[width=14cm]{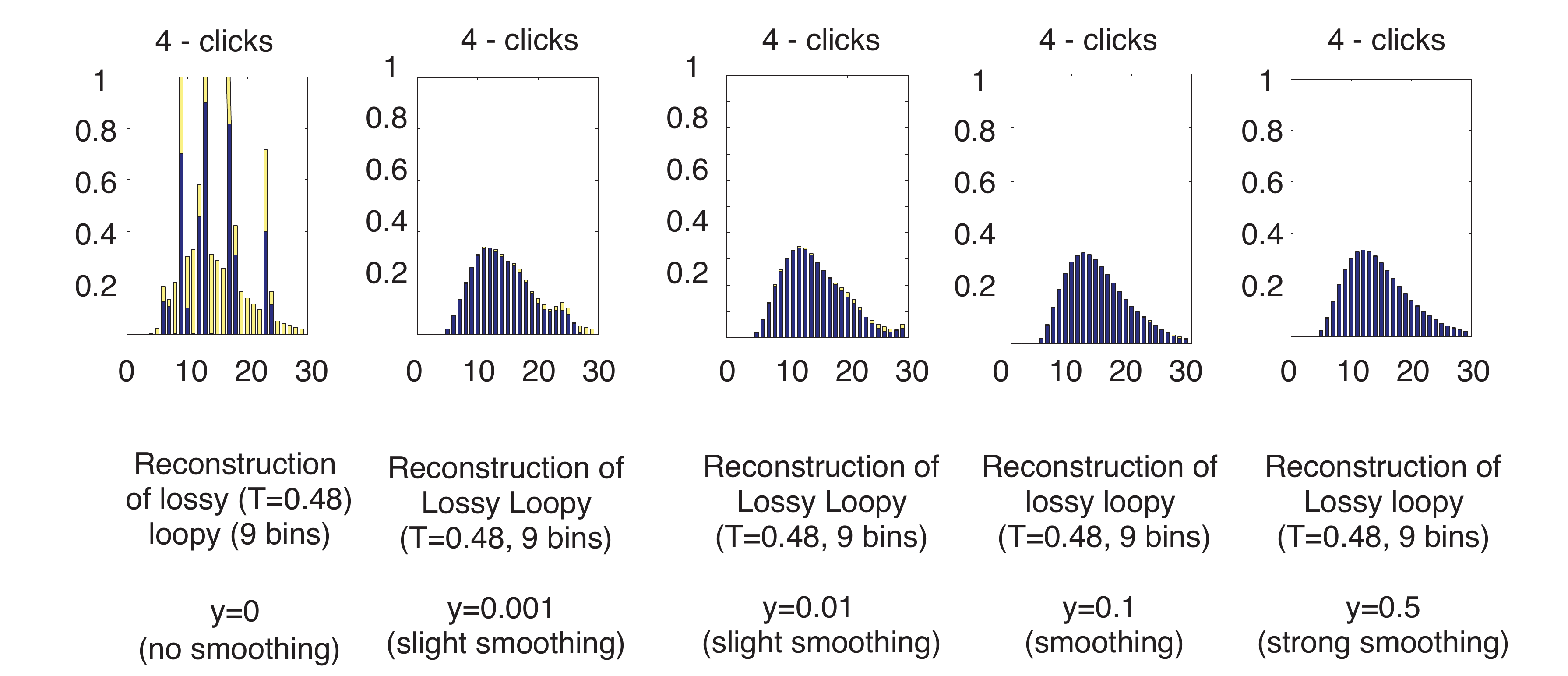}
\caption{Smoothing evolution for a lossy TMD detector (loss=52\%).  We
  show as an example the evolution of the POVM element $\pi_{4}
  = \sum_{i=0}^{60} \theta_i^{(4, {\rm rec})}
  \ket{i}\bra{i} $ as we increase the amount of smoothing (in $y$).  The
  yellow bars display $|\theta_i^{(4,{\rm theo})} - \theta_i^{(4, {\rm rec})}|$
  stacked on top of $\theta_i^{(4, {\rm rec})}$ . }
\label{lossy smoothing evolution}
\end{figure*}
\end{center} 
Fig.\ \ref{lossy smoothing evolution} presents the evolution of the
``\textsl{$4$ click}'' POVM element as we add more smoothing (or increase $y$
in Eq.\ (\ref{SDP-optim})). This element is chosen as an illustrative example
but more details can be found in Ref.\ \cite{Alvaro-Thesis}. The figure shows in
blue the coefficients $\theta_{i}^{(4)}$ in $\pi^{\mathrm{\mathrm{(rec)}}}_{4}
= \sum_{i=0}^{60} \theta_{i}^{(4)} |i\rangle\langle i|$, where
\textsl{\textrm{(rec)}} means reconstructed. In yellow, stacked on top of
$\theta_{i}^{(4)}$ we display $|\theta_{i}^{(4,\mathrm{rec})} - \theta
_{i}^{(4,\mathrm{theo})}|$, where \textsl{\textrm{theo}} refers to the
original POVM we used to generate the probability distribution. Clearly the
smoothing improves the result and the exact value of $y$ is rather
unimportant. A sharp feature that is preserved however is $\theta_{i}^{(4)} =0
$ for $i < 4$ proving a good agreement with the model.

\subsubsection{Loss-less TMD}
\begin{figure*}[ptbh]%
\includegraphics[width=12.5cm]{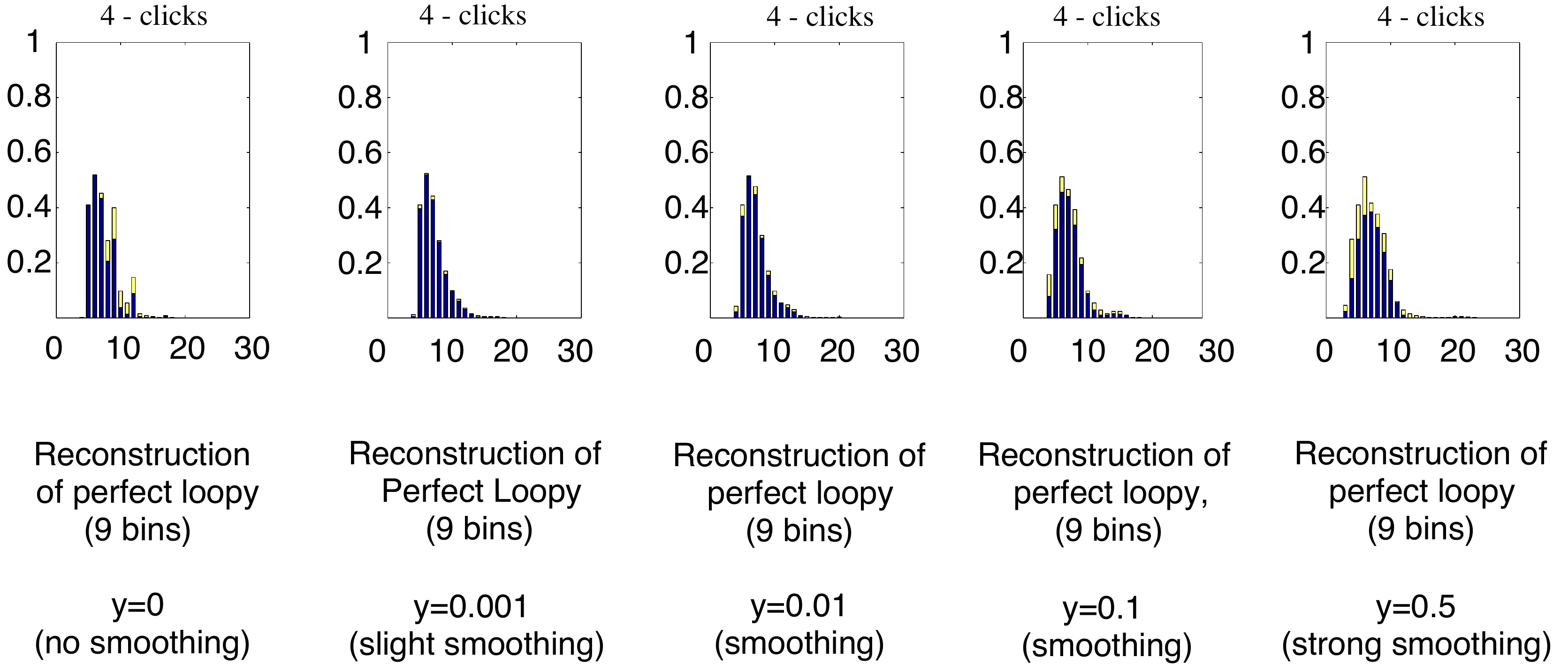}
\caption{Smoothing evolution for a perfect TMD detector (no loss). We show as
an example the evolution of the POVM element $\pi_{4}= \sum_{i=0}^{60}
\theta_{i}^{(4, \mathrm{rec})} |i\rangle\langle i| $ as we increase the amount
of smoothing (in $y$). The yellow bars display $|\theta_{i}^{(4,\mathrm{theo}%
)} - \theta_{i}^{(4, \mathrm{rec})}|$ stacked on top of $\theta_{i}^{(4,
\mathrm{rec})}$. }%
\label{loss-less loopy smoothing evolution}%
\end{figure*}
Fig.\ \ref{loss-less loopy smoothing evolution} shows also the ``\textsl{4
click}'' event and the error associated with the reconstruction (yellow). This
TMD shows in its distribution the finite number of bins as we described
earlier. The distribution is not as broad as that of the lossy-loopy and the
smoothing is therefore not so effective. The raw SDP, with $y=0$, performs
quite well, and the POVM is quite insensitive to the smoothing, although, when
given $1/2$ of the weight in the optimisation ($y=0.05$) the smoothing starts
to become harmful.

\subsubsection{Perfect number detector}
\begin{figure*}[th]%
\includegraphics[width=13cm]{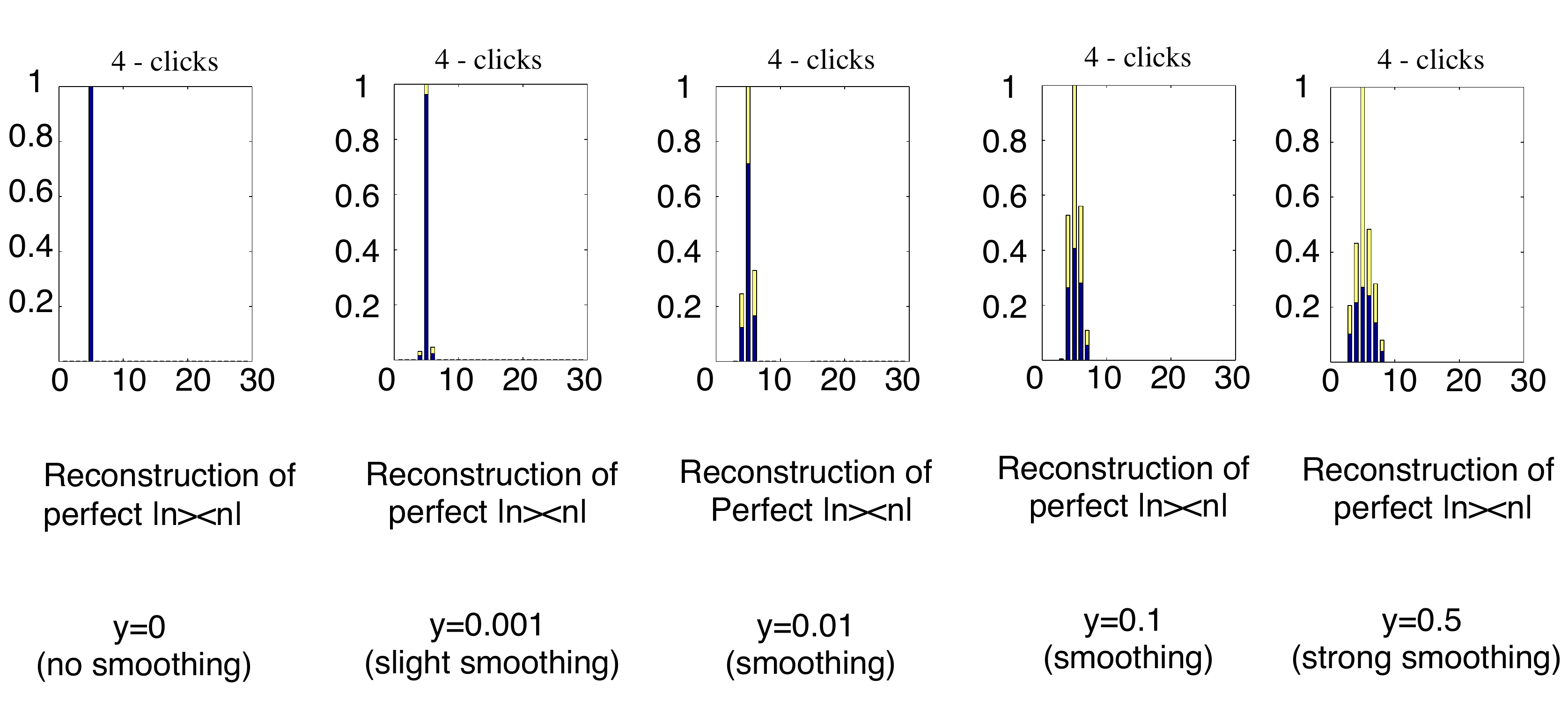}
\caption{Smoothing evolution for a perfect photon number detector, that is one
with $\pi_{n} = |n\rangle\langle n|$. We show as an example the evolution of
the POVM element $\pi_{4}= \sum_{i=0}^{60} \theta_{i}^{(4, \mathrm{rec})}
|i\rangle\langle i| $ as we increase the amount of smoothing (in $y$). The
yellow bars display $|\theta_{i}^{(4,\mathrm{theo})} - \theta_{i}^{(4,
\mathrm{rec})}|$ stacked on top of $\theta_{i}^{(4, \mathrm{rec})}$. }%
\label{n smoothing evolution}%
\end{figure*}
Fig.\ \ref{n smoothing evolution} shows also the ``\textsl{$4$ click}'' event
which in this case is simply $\pi_{4} = |4\rangle\langle4|$. A very
interesting feature is that the simple SDP with $y=0$ achieves a perfect
result. This happens in spite of using a mixed state as a probe state (mixture
of amplitudes $|\alpha|$ around $|\alpha\rangle\langle\alpha|$). The
reconstruction is then robust for very well defined and sharp features, where
the higher decaying coefficients do not introduce instabilities.

\subsubsection{Sharp POVM}
\begin{figure*}[ptbh]%
\includegraphics[width=12.5cm]{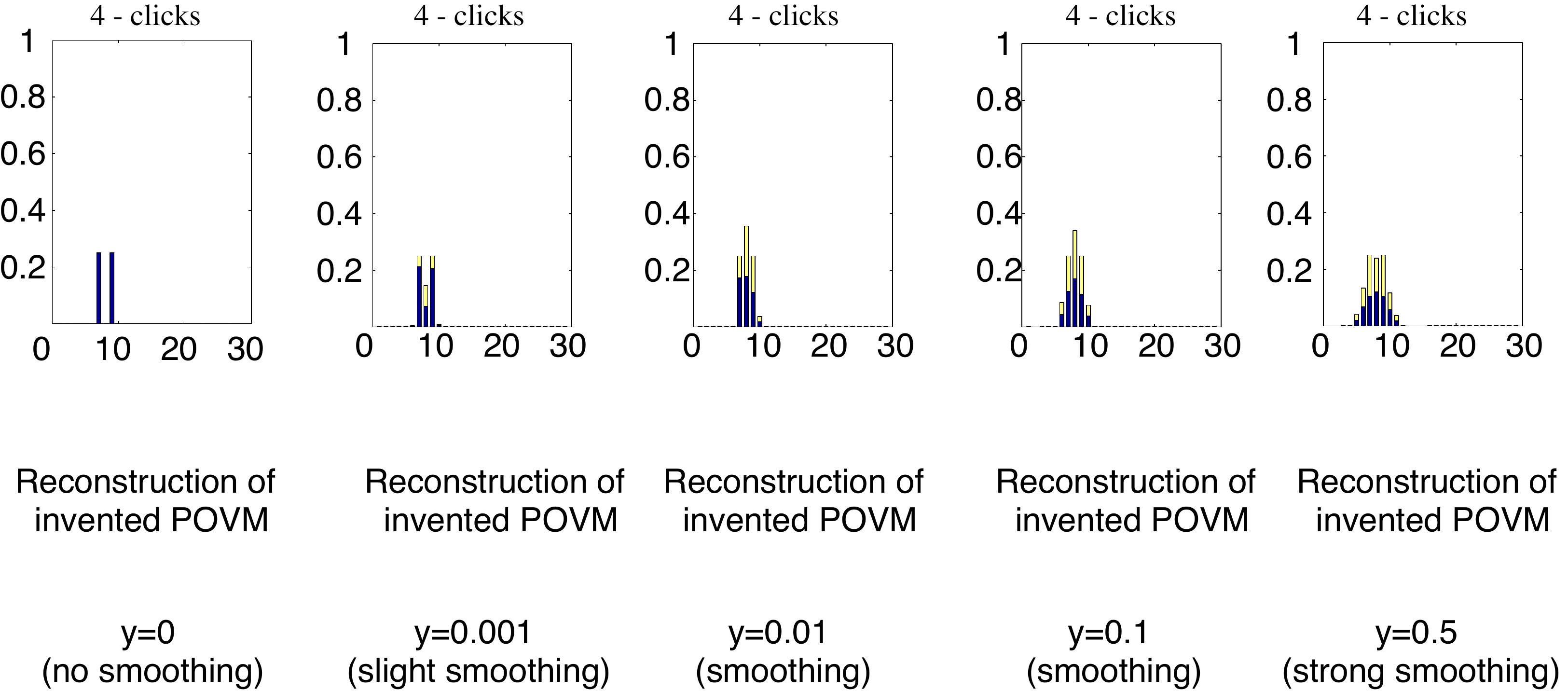} 
\caption{Smoothing evolution for an invented POVM with sharp variations.
Displayed is $\pi_{4} = |7\rangle\langle7| + |9\rangle\langle9|$. We show as
an example the evolution of the POVM element $\pi_{4}= \sum_{i=0}^{60}
\theta_{i}^{(4, \mathrm{rec})} |i\rangle\langle i| $ as we increase the amount
of smoothing (in $y$). The yellow bars display $|\theta_{i}^{(4,\mathrm{theo}%
)} - \theta_{i}^{(4, \mathrm{rec})}|$ stacked on top of $\theta_{i}^{(4,
\mathrm{rec})}$. }%
\label{sharp smoothing evolution}%
\end{figure*}
\begin{figure*}[ptbh]%
\includegraphics[width=12cm]{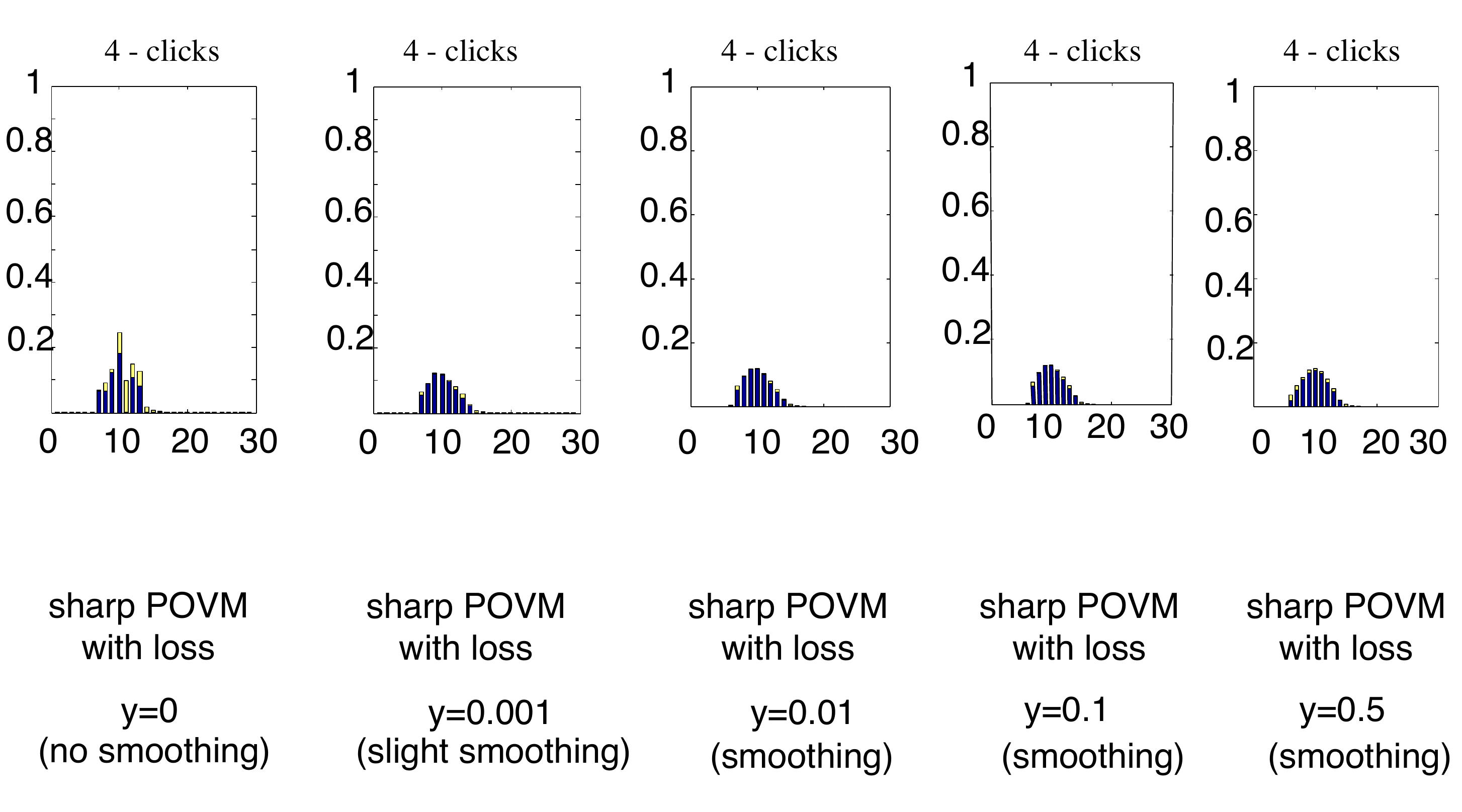} 
\caption{Smoothing evolution for an invented POVM with sharp variations which
has suffered a $20$\% loss (modelled by interposing a beam splitter)..
Displayed is the reconstruction of the element $\pi_{4} = |7\rangle\langle7| + |9\rangle\langle9|$
when it suffers the mentionned loss.
The reconstruction shows $\pi_{4}= \sum_{i=0}^{60}
\theta_{i}^{(4, \mathrm{rec})} |i\rangle\langle i| $ as we increase the amount
of smoothing (in $y$). The yellow bars display $|\theta_{i}^{(4,\mathrm{theo}%
)} - \theta_{i}^{(4, \mathrm{rec})}|$ stacked on top of $\theta_{i}^{(4,
\mathrm{rec})}$. }%
\label{sharp_loss smoothing evolution}%
\end{figure*}
We now discuss the situation of a POVM element that is not related to an
experiment, but has been artificially generated to identify the limit of the
smoothing regularisation. The element displayed in
Fig.\ \ref{sharp smoothing evolution} is $\pi_{4} = |7\rangle\langle7| +
|9\rangle\langle9|$ and we can see that $y=0.1$ is already too much smoothing.
Certainly to reconstruct a completely loss-less detector with such a structure
smoothing is not an appropriate strategy. We must remember however that all
current photon-number detectors that count particles do exhibit loss, and have
therefore some degree of smoothness in them.

\subsubsection{Sharp POVM with loss}

The previous case could have given the impression that the reconstruction fails
for a sharp POVM.  However one has to stress that smoothing (or a regularization
for the optimisation) is necessary when there is loss and an ill conditionned matrix
(which is a generic case in quantum optics using coherent states for detector tomography).
Therefore, it's worth considering what happens when an invented POVM (as the previous one)
is made more realistic adding some loss.  Fig.\ \ref{sharp_loss smoothing evolution} 
illustrates this, showing the reconstruction of the previous sharp POVM element which
has suffered a $20\%$ loss.  Indeed in this case we can see a clear improvement as the smoothing
helps regularize the optimisation.

\section{Conclusion}

As quantum information and computation implementations evolve, detectors are
becoming more complex. In addition, crypto security requires a careful
statement of assumptions idealy kept to a minimum. This, as we have seen,
calls for a black-box characterisation of the operators they implement. We
have seen the first implementation of this type of tomography. We discussed in
detail the first experimental realisation of quantum detector tomography
completing the triad of experimental state \cite{PhysRevA.40.2847,
PhysRevLett.70.1244, PhysRevA.60.674}, process \cite{chuang-1996,
PhysRevLett.78.390,PhysRevLett.80.5465,kwiat}, and detector tomography
\cite{NaturePhysTomo}. This detector characterisation opens up more flexible
and complex ways of detecting quantum states and accurately preparing
non-classical light.

The reconstruction methods are simple and efficient. However one has to pay
close attention to the subtleties behind the ill-conditioning of such
reconstructions whether its state or detector tomography. Fully characterising
a detector with this method can help get rid of complex or erroneous
assumptions in the modelling. Furthermore, once they are fully characterised,
one can re-design or alter the detectors with a direct feedback on their performance.

Detector tomography significantly benefits state tomography or
metrology, as well as state preparation and the implementation of
protocols in quantum information requiring detectors in state
manipulation. Importantly, it enables the use of detectors that are
noisy, non-linear or that operate outside their intended range. One
conclusion is that lossy detectors are often just as useful as perfect
ones, as long as we know the exact POVMs and one can describe the rest
of our experiment accordingly.  This method will also allow the
benchmarking of similar detectors making performance comparisons
possible.  Indeed one can also ask question concerning the power each
detector has for preparing non-classical states.  This opens a path
for the experimental study of novel concepts such as the
non-classicality of detectors. Another promising avenue is to
translate  homodyne
tomography techniques to optical detector tomography.  For example
defining the detector tomography equivalents of balanced
noise-reduction, direct measurement of the Wigner function or pattern
functions).  Naturally an immediate next step would involve
characterizing detectors with off diagonal terms and phase
sensitivity. 

\subsubsection*{Acknowledgements}

This work has been supported by the EU (COMPAS, HIP, MINOS, Marie Curie, and QAP), 
the EPSRC, The Heinz-Durr Stipendienprogram of the SDV, the QIP-IRC, Microsoft Research, the EURYI Award Scheme 
and the Royal Society.

\end{document}